\documentclass[%
 reprint,
 superscriptaddress,
 amsmath,amssymb,
 aps,
prb,
]{revtex4-2}

\usepackage{graphicx}% Include figure files
\usepackage{dcolumn}% Align table columns on decimal point
\usepackage{bm}% bold math
\usepackage{bbold}% bold math
\usepackage{mathtools}% math
\usepackage{hyperref}  
\hypersetup{
    colorlinks=true,
    linkcolor=blue,
    filecolor=blue,      
    urlcolor=blue,
    citecolor=blue,
    }
\usepackage{xcolor}
\usepackage{caption}% bold math
\newcommand{\kk}{\boldsymbol{k}}
\newcolumntype{C}{>{$\displaystyle}c<{$}}

\makeatletter
\def\maketitle{
\@author@finish
\title@column\titleblock@produce
\suppressfloats[t]}
\makeatother

\begin{document}

\preprint{APS/123-QED}

\title{Nonflat bands and chiral symmetry in magic-angle twisted bilayer graphene}

\author{Miguel S\'anchez S\'anchez}
\email{miguel.sanchez@csic.es}
\affiliation{Insituto de Ciencia de Materiales de Madrid ICMM-CSIC. Madrid (Spain)}
\author{Jos\'e Gonz\'alez}
\affiliation{Insituto de Estructura de la Materia IEM-CSIC. Madrid (Spain)}
\author{Tobias Stauber}
\email{tobias.stauber@csic.es}
\affiliation{Insituto de Ciencia de Materiales de Madrid ICMM-CSIC. Madrid (Spain)}

% \date{\today}

\begin{abstract}
In this work, we study an interacting tight-binding model of magic-angle twisted bilayer graphene (MATBG), with a twist angle of $1.05^\circ$. We derive effective theories based on a mean-field normal state at charge neutrality, thereby including the renormalizations coming from integrating out high-energy modes. In these theories, the flat bands display a sizable increase of the bandwidth, suggesting the renormalization of the magic angle. Additionally, the corresponding wavefunctions flow towards the limit of perfect particle-hole symmetry and sublattice polarization (the 'chiral' limit). We further represent the flat bands in the 'vortex Chern' basis and discuss the implications on the dynamics, regarding the 'flat' and 'chiral' symmetries of MATBG, as manifested in the symmetry-broken states at neutrality.    

\end{abstract}

%\keywords{Suggested keywords}%Use showkeys class option if keyword
                              %display desired
\maketitle

%\tableofcontents

% \paragraph*{\color{blue}{Introduction.}}
\section{Introduction}

\begin{figure}[t]
     \centering
     \includegraphics[width=1\linewidth]{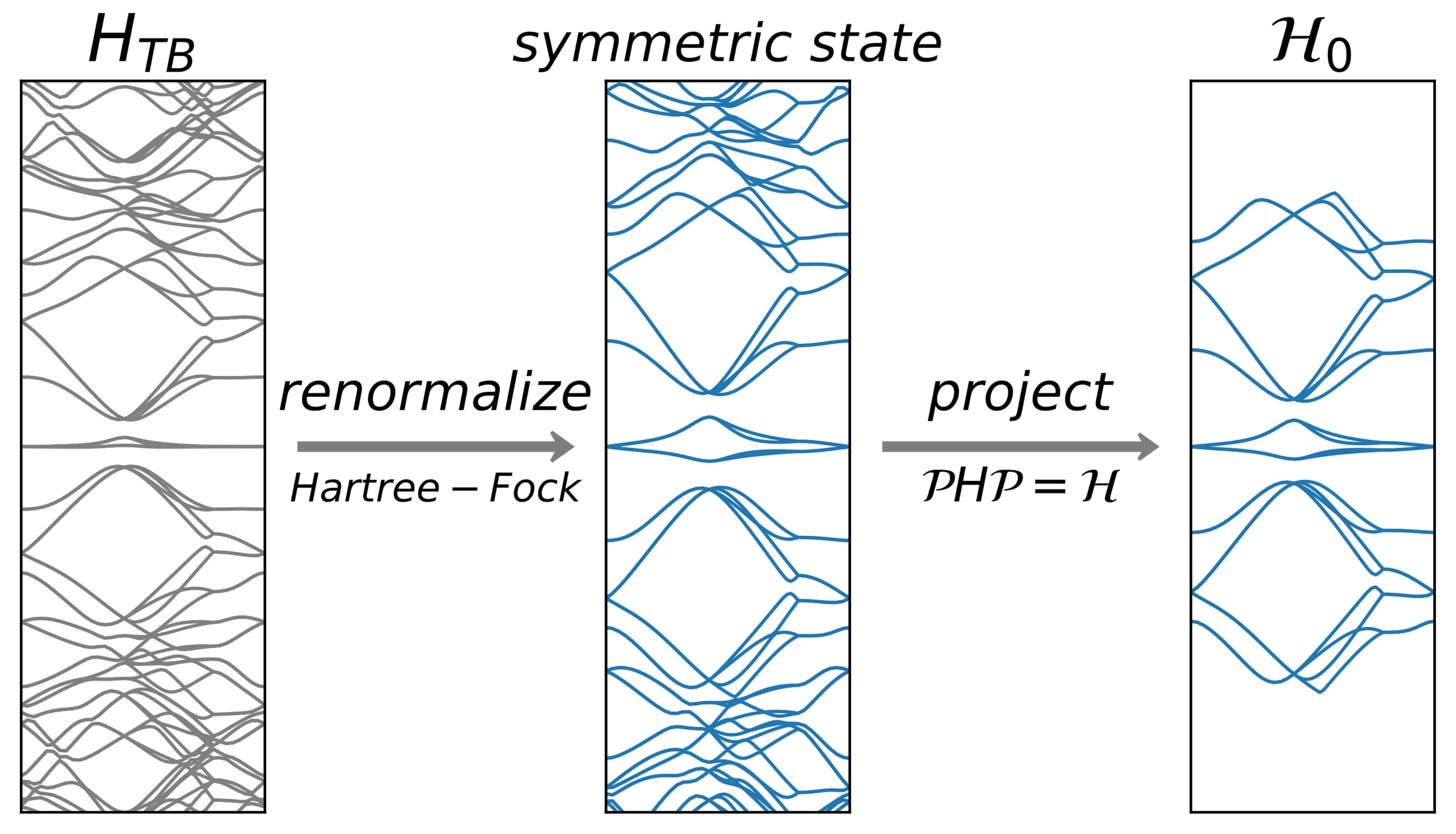}     
     \caption{Many-body projection scheme. First, we obtain a renormalized, self-consistent, symmetry-preserving state. We then project onto excitations of the central bands, resulting in a renormalized dispersion, $\mathcal{H}_0$, with nonflat bands.} 
     \label{fig1}
\end{figure}

Magic-angle twisted bilayer graphene (MATBG), as a type of moir\'e material \cite{Mak2022,Nuckolls2024}, hosts a mesoscopic unit cell and flat electronic bands at charge neutrality. It represents a tunable platform for the study of strong interactions and topology \cite{tarnopolski19,liu19,song21}, as well as their interplay, thus attracting great attention in the recent years. 
% Experimental and theoretical efforts have unveiled a rich phase diagram, exhibiting in a single material most of the known phenomena in Condensed Matter Physics like correlated insulators \cite{Bultinck20, lian21, xie20, kang19,Lu2019,Saito2020,Stepanov2020} Integer \cite{wang24,Yu2022,Nuckolls2020,stepanov21,Sharpe2019,Pierce2021,Das2021} and Fractional \cite{Xie2021,ledwith20,Reppellin20} Chern insulators, Kondo physics \cite{Wong2020,Zondiner2020,rai24,Datta2023,hu23_2,hu23,zhou24} and Superconductivity\cite{Cao2018,Cao2021,Lu2019,Saito2020,Stepanov2020,Oh2021,Liu2021,Arora2020}, among others \cite{Saito2021,Rozen2021,cao2020,Jaoui2022}.
Experimental and theoretical efforts have unveiled a rich phase diagram, exhibiting in a single material phenomena like correlated states \cite{Bultinck20, lian21, bernevig21_2, xie20, liu21, parker21, soejima20, Hofmann22,wagner22, wang2024electronphononcouplingtopological,wang24,Lu2019,Saito2020,Stepanov2020,Yu2022,Nuckolls2020,stepanov21,Sharpe2019,Pierce2021,Das2021,Zhang2022,Kerelsky2019,adhikari24}, Kondo physics \cite{Wong2020,Zondiner2020,rai24,Datta2023,hu23_2,hu23,zhou24} and superconductivity \cite{Cao2018,Cao2021,Lu2019,Saito2020,Stepanov2020,Oh2021,Liu2021,Arora2020,gonzalez19,wang24_2,Christos2023,ingham2023quadraticdiracfermionscompetition,Cea2021,Long2024,sanchez24_2}, among others \cite{Xie2021,Saito2021,Rozen2021,cao2020,Jaoui2022,Andrews20}.

The large number of atoms in the moir\'e unit cell, and consequently large number of electronic bands, calls for the construction of effective theories of MATBG with a small number of degrees of freedom, that ought to capture accurately the dispersion, symmetry and topology of the low-energy bands. Among such theories there are continuum models \cite{carr19,fang2019angledependentitabinitio,kang23,koshino20}, including the seminal Bistritzer-MacDonald (BM) model \cite{bistritzer11}, Wannier orbital \cite{carr19_2,koshino18,vafek18,po19,calderon20,vafek21,cao21,bennett24,yuan18}, momentum space \cite{ledwith2024nonlocalmomentschernbands,miao23,ingham2023quadraticdiracfermionscompetition} and the topological heavy fermion (THF) model \cite{song22}. 

Their very small bandwidth renders the flat bands prone to distortions coming from otherwise minor perturbations, like lattice relaxation \cite{kang23,fang2019angledependentitabinitio,herzogarbeitman2024heavyfermionsefficientrepresentation,Yoo2019} or strain \cite{herzogarbeitman2024heavyfermionsefficientrepresentation,bi19,Kazmierczak2021}. In this respect, an often overlooked piece of the physics is the interacting effects of the high-energy degrees of freedom that are removed in the effective theory, but remain implicit in the ground state.
% It is usually implicit that the 'active' degrees of freedom, say, the Wannier orbitals in multi-orbital models, the flat-band wavefunctions in projected models or the $f$ fermions in the THF model, are decoupled from the integrated-out \cite{Bultinck20,liu21,parker21,soejima20,lian21,Hofmann22,wang23,wagner22,song22}.
We argue that virtual processes between the 'remote' (high-energy) and 'active' (low-energy) modes are significant and must be accounted for \cite{Potasz21}, at least approximately. An earlier work addressed this problem using a Renormalization Group approach \cite{vafek20}.

In this Article, we consider twisted bilayer graphene with a twist of $1.05^\circ$, around the nominal magic angle \cite{kang23}. We obtain a normal state at charge neutrality that preserves the symmetries and incorporates the Hartree-Fock potentials of the occupied graphene $\pi$ orbitals self-consistently; any low-energy theory of MATBG ought to reproduce the dispersion and wavefunctions of the normal, rather than the non-interacting, state. We construct effective models for MATBG using a many-body projection scheme onto the central bands of this symmetric state, sketched in Fig. \ref{fig1}.

In the renormalized system the dispersion widens drastically with the flat bandwidth reaching several times the non-interacting one, becoming comparable to the energy scale of the Coulomb interactions among flat electrons. This suggests the tuning of the magic angle (or magic 'range' \cite{carr19}) by the interactions, in a similar way to lattice relaxation \cite{nam17,carr19}. 
Furthermore, the flat-band Hilbert space flows towards the limit of perfect particle-hole symmetry and sublattice polarization (the chiral limit \cite{tarnopolski19,ledwith20,vafek20}).

At the same time, there are important modifications of the symmetry of MATBG \cite{Bultinck20,bernevig21,song22}. The $U(4)_\text{flat}$ symmetry of the 'nonchiral-flat' limit, that relies on the existence of very flat bands, is strongly broken; instead, the 'chiral-nonflat' $U(4)_{\text{chiral}}$ group prevails as an approximate symmetry independently of the bandwidth. Such properties become apparent in the symmetry-broken phases at charge neutrality. 
% Lastly, we comment on the experimental imprints of our results on the recently observed spiral states \cite{Nuckolls2023}.

\begin{figure}[t]
     \centering
     \includegraphics[width=1\linewidth]{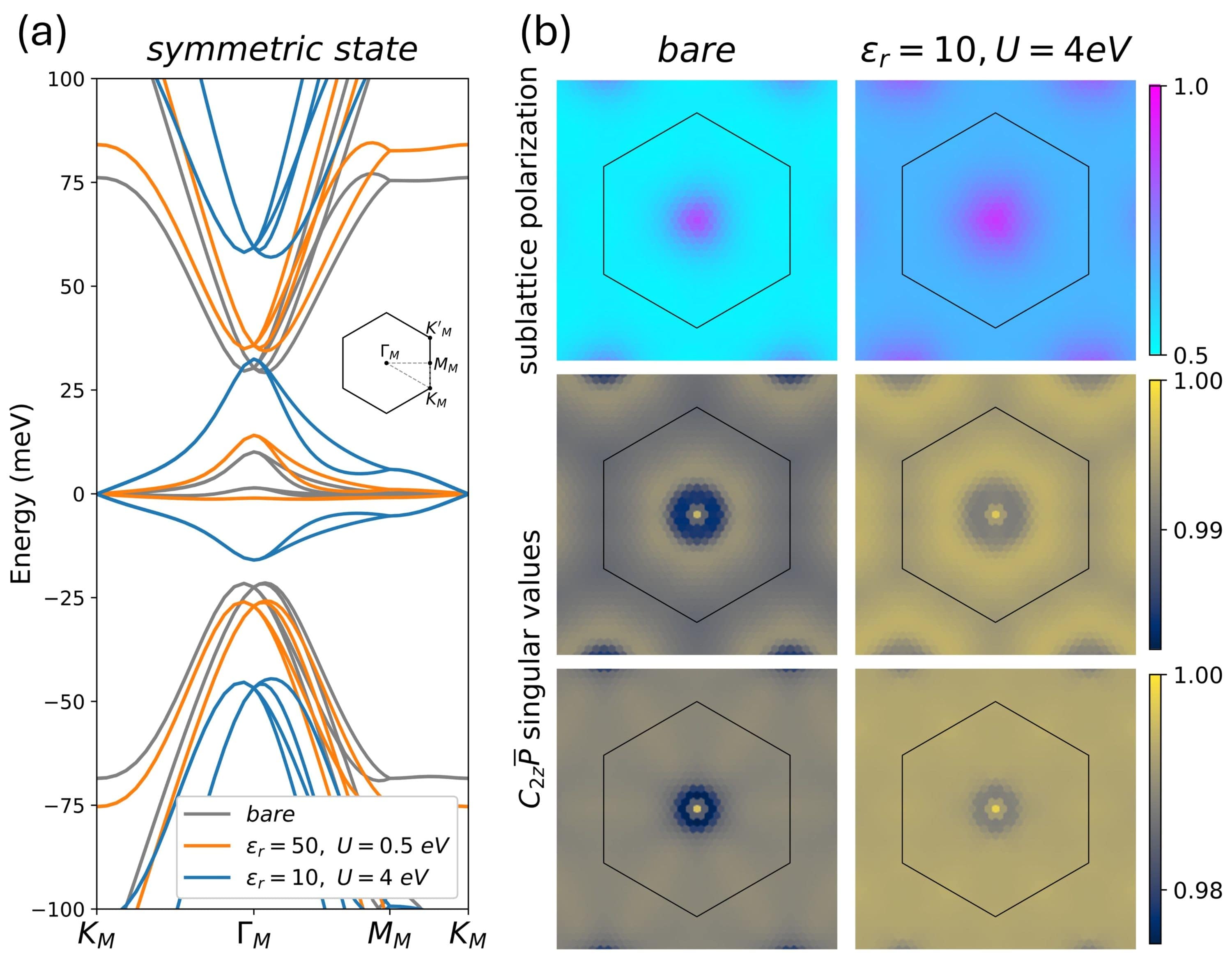}
     \caption{(a) Band structure of the symmetric state for different interaction strengths, compared to the non-interacting (bare) bands. The inset shows the Brillouin zone and the momentum lines plotted. (b) Hilbert space of the bare and renormalized ($\epsilon_r=10,$ $U=4$ eV) flat bands. In the top row we show the positive eigenvalue of the sublattice operator, $\overline{\sigma}_z$, (sublattice polarization) as a function of momentum. The remaining eigenvalues are obtained from $180 ^\circ$ rotations and sign changes. In the bottom rows we plot the two distinct singular values of the projected particle-hole operator $C_{2z}\overline{P}$, which are degenerate in pairs.} 
     \label{fig2}
\end{figure}

% \paragraph*{\color{blue}{Symmetric state and many-body projection.}}
\section{Symmetric state and many-body projection}

We start with a tight-binding model for MATBG with one orbital per carbon atom; there are 11908 atoms in the unit cell of the $1.05^\circ$ twisted structure. In-plane atomic relaxations are included following Ref. \cite{nam17}. The crystalline symmetries of the lattice include six-fold rotations around the $z$ axis, $C_{6z}$, and two-fold rotations around the $y$ axis between the layers, $C_{2y}$; below we will be addressing the products $C_{3z} = C_{6z}^2$, $C_{2z} = C_{6z}^3$ and $C_{2x} = C_{2y}C_{2z}$ also. Spinless time-reversal, $\mathcal{T}$, acting as complex conjugation, is a symmetry if we neglect spin-orbit coupling.

We include the electron-electron interactions through the Coulomb potential, screened by metallic plates at distances $\pm\xi/2 = \pm 5$ nm from the material, $V(\boldsymbol{r}-\boldsymbol{r'}) = \frac{e^2}{4\pi \epsilon_0 \epsilon_r}\sum_{n=-\infty}^\infty \frac{(-1)^n}{||\boldsymbol{r} - \boldsymbol{r'} + n\xi \boldsymbol{\hat{z}}||}$. The Hubbard energy $U$ regularizes the Coulomb potential at $\boldsymbol{r}=0$, and $\epsilon_r$ includes the effect of the substrate and the self-screening \cite{Pizarro19,vanhala20}. For reference, the numerical value of $e^2/4\pi \epsilon_0$ is $14.4$ eV$\times \text{\AA}$

The total Hamiltonian, ${H}$, consists of the tight-binding part and the interaction, ${H} = H_{\text{TB}} + {H}_{\text{int}}$, with
\begin{align}
    H_{\text{TB}} =& \sum_{\boldsymbol{r}\boldsymbol{r'}s} t(\boldsymbol{r}-\boldsymbol{r'}) c^\dagger_{\boldsymbol{r}s}c_{\boldsymbol{r'}s}, \nonumber \\ 
    % H_{\text{int}} =& \frac{1}{2} \smashoperator[r]{\sum_{\boldsymbol{r}\neq \boldsymbol{r'},ss'}} V(\boldsymbol{r}- \boldsymbol{r'}) \delta n_{\boldsymbol{r}s} \delta n_{\boldsymbol{r'}s'} + \sum_{\boldsymbol{r}} U \delta n_{\boldsymbol{r}\uparrow} \delta n_{\boldsymbol{r}\downarrow}.
    H_{\text{int}} =& \frac{1}{2} \sum_{\boldsymbol{r}\neq \boldsymbol{r'},ss'} V(\boldsymbol{r}- \boldsymbol{r'}) \delta n_{\boldsymbol{r}s} \delta n_{\boldsymbol{r'}s'} 
    + \sum_{\boldsymbol{r}} U \delta n_{\boldsymbol{r}\uparrow} \delta n_{\boldsymbol{r}\downarrow}.
\label{hamiltonian}
\end{align}

Here, $c^\dagger_{\boldsymbol{r}s}$($c_{\boldsymbol{r}s}$) is the creation (annihilation) operator of an electron with spin $s$ at position $\boldsymbol{r}$ and $t(\boldsymbol{r})$ denotes the hopping function \cite{laissandiere10,moon12}; $\delta n_{\boldsymbol{r}s} = n_{\boldsymbol{r}s} - \frac{1}{2} =  c^\dagger_{\boldsymbol{r}s}c_{\boldsymbol{r}s} - \frac{1}{2}$ is the spin-$s$ density at position $\boldsymbol{r}$ relative to a uniform background.
 
We find a self-consistent solution of ${H}$ in mean-field (Hartree-Fock) theory that preserves spin rotations, time-reversal symmetry and the crystallographic symmetries (see Appendix \ref{appb} for details).
% (\footnote{see Supplementary Material, which includes the references \cite{dossantos12,nam17,slater54,laissandiere10,moon12,Giuliani_Vignale_2005,Kudin2002,bernevig21,Bultinck20,lopez20,liu19,song21,song19,kang23,herzogarbeitman2024heavyfermionsefficientrepresentation,xie2024chernbandsoptimallylocalized,gunawardana24,song22,blason22,shi2024moireopticalphononsdancing,wang2024electronphononcouplingtopological,wang24_2,angeli19,kwan2023electronphononcouplingcompetingkekule,sánchez2023correlatedinsulatorsmagicangle,Chatterjee22,sanchez24,alexandrinata14,carr19,po19}} for details). 
In Fig. \ref{fig2}(a) we plot the band structures of the symmetric state with $\epsilon_r=10$, $U=4$ eV and $\epsilon_r=50$, $U=0.5$ eV (the weak coupling limit corresponds to the physical situation if the self-screening is strong \cite{gonzalez21}), and compare them with the bare band structure. The flat bandwidth increases approximately linearly with $\epsilon_r^{-1}$, growing from $8.7$ meV to $15.3$ meV to $48.4$ meV in the bare, $\epsilon_r=50$ and $\epsilon_r=10$ bands, respectively. Another relevant quantity is the gap between the flat and remote bands, controlling the validity of the flat-band projection \cite{ledwith2024nonlocalmomentschernbands}; the gaps increase from $20.2$ to $26.8$ meV on the electron side and from $23.9$ to $30.8$ meV on the hole side in the bare and $\epsilon_r=10$ bands, respectively. 

Not only the energies, but also the wave functions are renormalized. In Fig. \ref{fig2}(b) we study the properties of Hilbert space of the flat bands; we plot the sublattice polarization, this is, the eigenvalues of the sublattice operator projected to the flat bands, $\overline{\sigma}_z$, and the singular values of the 'particle-hole' operator \cite{song19,song21,kang23,herzogarbeitman2024heavyfermionsefficientrepresentation} projected to the flat bands, $C_{2z}\overline{P}$. The system flows towards a perfectly sublattice polarized and particle-hole symmetric limit with eigenvalues, or singular values, equal to $1$. The increase of the $\overline{\sigma}_z$ eigenvalues, i.e. the flow towards the chiral limit \cite{tarnopolski19,ledwith20}, was already noted in Ref. \cite{vafek20}.

Using the projector onto many-body excitations of the lowest $n_B$ bands (per spin) of the normal state, $\mathcal{P}$, we can obtain an effective model for MATBG with the Hamitonian ${\mathcal{H}} = \mathcal{P}H\mathcal{P} = \mathcal{H}_0 + {\mathcal{H}}_{\text{int}}$,
\begin{align}
    &\mathcal{H}_0 = \smashoperator[r]{\sum_{\boldsymbol{k} nm s}} {h}(\boldsymbol{k})_{nm} c^\dagger_{\kk n s} c_{\kk m s}, \nonumber \\
    % &{\mathcal{H}}_{\text{int}} = \frac{1}{2} \smashoperator[r]{\sum_{\boldsymbol{r}\neq \boldsymbol{r'}, ss'}} V(\boldsymbol{r}- \boldsymbol{r'}) \delta{\rho}_{\boldsymbol{r}s} \delta{\rho}_{\boldsymbol{r'}s'} + \sum_{\boldsymbol{r}} U \delta{\rho}_{\boldsymbol{r}\uparrow} \delta{\rho}_{\boldsymbol{r}\downarrow}.
     &{\mathcal{H}}_{\text{int}} = \frac{1}{2} {\sum_{\boldsymbol{r}\neq \boldsymbol{r'}, ss'}} V(\boldsymbol{r}- \boldsymbol{r'}) \delta{\rho}_{\boldsymbol{r}s} \delta{\rho}_{\boldsymbol{r'}s'} + \sum_{\boldsymbol{r}} U \delta{\rho}_{\boldsymbol{r}\uparrow} \delta{\rho}_{\boldsymbol{r}\downarrow}.
\end{align}
In this expression, the operator $c^\dagger_{\kk n s}(c_{\kk n s})$ creates (annihilates) an electron on the $n^{\text{th}}$ mean-field band with momentum $\kk$ in the first Brillouin zone and spin $s$, and $n,m$ are restricted to the central $n_B$ bands, $n,m = \{-n_B/2,...,n_B/2\}$. We take $n_B=4$ and $20$, for which the band projector is well-defined, see Appendix \ref{appb} for details. 
%In both cases, the $n_B$ lowest bands are gapped from the remaining bands at all $\kk$, so the projector is well-defined. 

The matrix ${h}(\kk)_{nm}$ contains the dispersion of the normal state plus a counter-term to avoid double-counting the valence bands $0<n\leq -n_B/2$, and a term related to the order of the fermionic operators, see Appendix \ref{appb} for details. On the other hand, ${\rho}_{\boldsymbol{r}s}$ is the density relative to charge neutrality,
\begin{gather}
    \delta{\rho}_{\boldsymbol{r}s} = {\rho}_{\boldsymbol{r}s} - {\rho}_{0 \boldsymbol{r}}, \quad \quad {\rho}_{\boldsymbol{r}s} = \sum_{\mathclap{\kk \boldsymbol{q} n m}} \langle \kk n | \boldsymbol{r} \rangle \langle \boldsymbol{r} | \boldsymbol{q} m \rangle c^\dagger_{\kk n s} c_{\boldsymbol{q} m s}, \nonumber \\
    {\rho}_{0\boldsymbol{r}} = \sum_{\mathclap{\kk,n<0}} |\langle \boldsymbol{r} | \kk n\rangle|^2  \approx \frac{1}{2}\sum_{\kk n} | \langle \boldsymbol{r} | \kk n\rangle|^2 , 
\label{formfactors}
\end{gather}
where $\langle \boldsymbol{r} | \kk n \rangle$ is the wave function amplitude of mode $c^\dagger_{\kk n s}$ at position $\boldsymbol{r}$; again $n,m=\{-n_B/2,...,n_B/2\}$. The choice of $\rho_{0\boldsymbol{r}}$ is consistent with the approximate particle-hole symmetry.

\begin{figure}[t]
     \centering
     \includegraphics[width=1\linewidth]{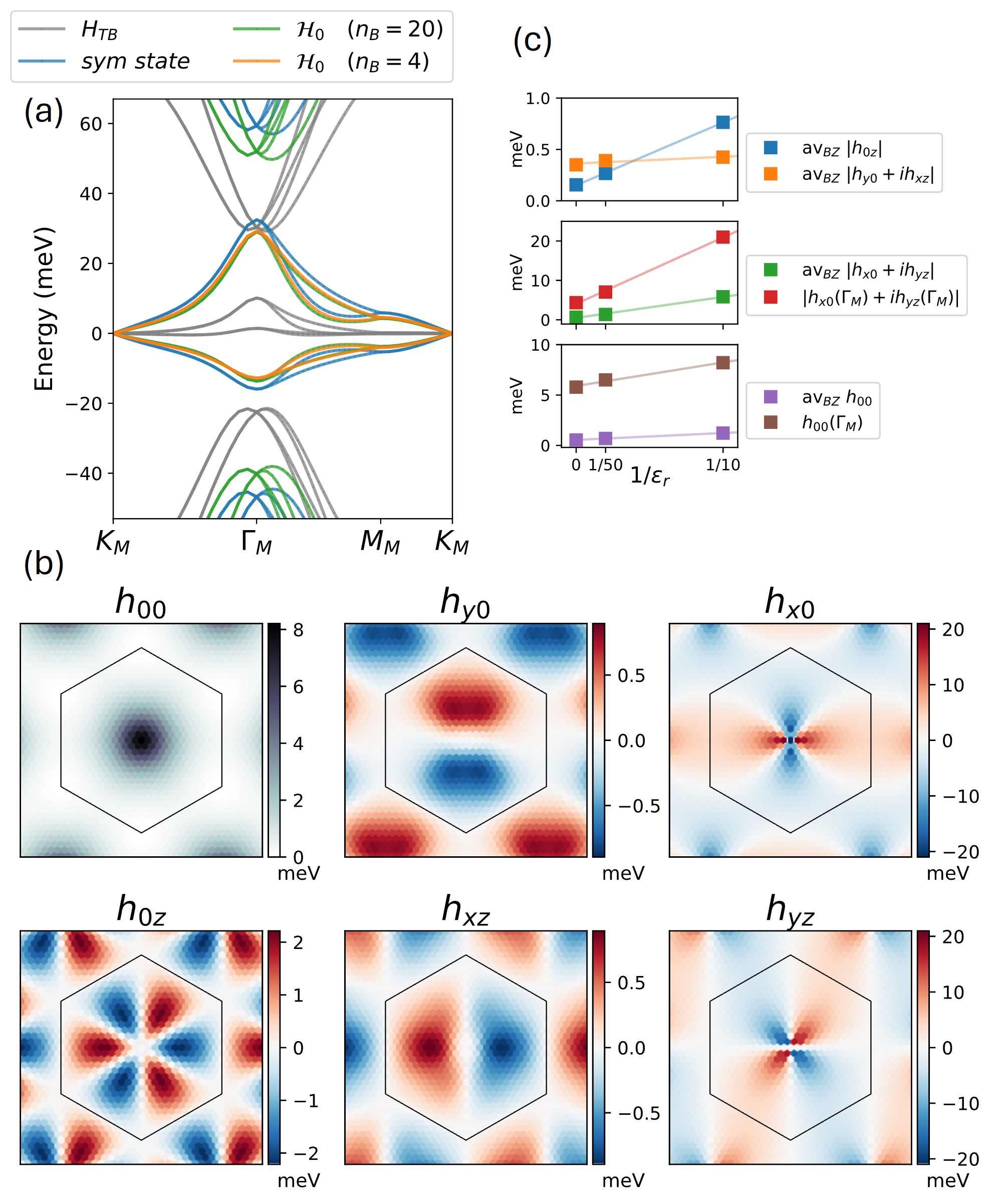}
     \caption{(a) Comparison of the bare bands, the bands of the symmetric state and the bands of the projected models with $n_B=4$ and $20$. (b) The nonzero components of ${h}(\boldsymbol{k})$ in the vortex Chern basis for $n_B=4$. Notice the different scale bars  ($\epsilon_r=10$, $U=4$ eV in (a-b)). (c) The components of ${h}(\boldsymbol{k})$ as a function of $1/\epsilon_r$; data points for $\epsilon_r^{-1}=0$, $U=0$ eV, $\epsilon_r=50$, $U=0.5$ eV and $\epsilon_r=10$, $U=4$ eV. The average over the Brillouin zone is denoted av$_{{BZ}}$ and the lines are linear fits.}
     \label{fig3}
\end{figure}

The band structure of $\mathcal{H}_0$ is shown in Fig. \ref{fig3}(a) with $\epsilon_r =10$, $U=4$ eV, for $n_B=4$ and $20$. In both cases, the dispersion is only slightly narrower than in the normal state, evidencing the crucial effects coming from the removed bands in the projected theories.

\section{Vortex Chern gauge and nonflat bands}
% \paragraph*{\color{blue}{Vortex Chern gauge and nonflat bands.}}

In the flat-band ($n_B=4$) theory, we obtain a valley and sublattice polarized basis by diagonalizing the projeccted valley ($\overline{\tau}_z$) \cite{lopez20} and sublattice ($\overline{\sigma}_z$) operators. We use $\tau_i$ and $\sigma_i$ to denote the identity ($i=0$) and Pauli ($i=x,y,z$), matrices in valley and sublattice space, valley $K(K')$ and sublattice $A(B)$ having eigenvalue $1(-1)$ under $\tau_z$ and $\sigma_z$, respectively. The resulting bands are topological with Chern numbers equal to $\sigma_z\tau_z$ \cite{Bultinck20,liu19,song21}. 

The 'Chern gauge' \cite{Bultinck20,bernevig21} fixes at each $\kk$ the relative phase between bands with the same Chern number, leaving an ambiguity of the form $\text{exp}(i\phi(\kk)\sigma_z\tau_z)$ \cite{Bultinck20,kwan21,Hofmann22} (see Appendix \ref{appc} for details on the gauge fixing and symmetry representations). We fix $\phi(\kk)$ by requiring that the Wannier functions obtained from the Chern bands are maximally localized \cite{xie2024chernbandsoptimallylocalized,gunawardana24}, following the algorithm of Ref. \cite{xie2024chernbandsoptimallylocalized}; the Bloch states are periodic and smooth except at the $\Gamma_M$ point, where the Berry connection acquires a vortex with a winding number equal to the Chern number. This 'vortex Chern gauge' is defined up to a global sign of all wave functions.

The renormalized dispersion can now be decomposed into Hermitian components,
\begin{align}
    \mathcal{H}_0 = \smashoperator[r]{\sum_{\kk}} c^\dagger_{\kk} {h}(\kk) c_{\kk}, \quad \quad
    {h}(\kk) = \smashoperator[r]{\sum_{ij}} {h}_{ij}(\kk) \sigma_i \tau_j,
\end{align}
where $c^\dagger_{\kk} = c^\dagger_{\kk \alpha}$ is a vector of creation operators and ${h}(\kk) = {h}(\kk)_{\alpha \beta}$ a matrix, $\alpha, \beta$ denoting the valley-sublattice-spin flavors and $i,j = 0,x,y,z$. By construction ${h}(\kk)$ is proportional to the unit spin matrix, and by $C_{2z}\mathcal{T}$ and valley conservation symmetries 
% enforce $\big[\sigma_x \mathcal{K},\overline{h}(\kk)\big] = \big[\tau_z, \overline{h}(\kk) \big] = 0$, where $\mathcal{K}$ is the complex conjugation. Consequently, 
only the coefficients ${h}_{00}(\kk)$, ${h}_{0z}(\kk)$, ${h}_{x0}(\kk)$, ${h}_{yz}(\kk)$, ${h}_{y0}(\kk)$ and ${h}_{xz}(\kk)$ are nonzero. 

In Fig. \ref{fig3}(b) we show the non-vanishing coefficients for $\epsilon_r=10$, $U=4$ eV. The dominant terms are ${h}_{x0}$ and ${h}_{yz}$, ${h}_{00}$ is also sizable at $\Gamma_M$. The 'minimal' Hamiltonian with ${h}(\kk) = {h}_{x0}(\kk) \sigma_x + {h}_{yz}(\kk) \sigma_y \tau_z$ inherits the Dirac cones at $K,K'$ and has a bandwidth of $41.9$ meV, to be compared with the bare bandwidth of $8.7$ meV. Furthermore, the representation $\text{exp}(-2\pi i/3 \ \sigma_z\tau_z)$ of $C_{3z}$ forces the phase of ${h}_{x0}(\kk) + i {h}_{yz}(\kk)$ to wind $2$ mod $3$ times around $\Gamma_M$. Band topology manifests itself at $\Gamma_M$ where the representation of $C_{3z}$ changes to $\mathbb{1}$, hence ${h}_{x0}(\Gamma_M) + i{h}_{yz}(\Gamma_M)$ need not vanish and there is a vortex in ${h}_{x0}(\kk) + i{h}_{yz}(\kk)$.

In Fig. \ref{fig3}(c) we summarize the main features of ${h(\kk)}$ as a function of $1/\epsilon_r$
% , with data points for the bare system ($\epsilon_r \to \infty$ and $U=0$ eV), for $\epsilon_r=10$ and $U=4$ eV, and for $\epsilon_r=50$ and $U=0.5$ eV. 
The different components grow linearly with $1/\epsilon_r$, ${h}_{x0}$ and ${h}_{yz}$ becoming the largest with increasing coupling, already prevailing at $\epsilon_r \lesssim 50$. 
% For $\epsilon_r=50$ the renormalized and bare bandwidths are still comparable, see also Fig. 
% On the other hand, $\overline{h}_{y0}$ and $\overline{h}_{xz}$ increase only very weakly remaining small.

\begin{table}
    \centering
    \renewcommand{\arraystretch}{1.25} % Default value: 1
    \begin{tabular}{ccc}
        \hline \hline
          & \multicolumn{2}{c}{$E - E_\text{SYM}$ (meV)} \\
        % \hline
         & $\epsilon_r=10$, $U=4$ eV & $\epsilon_r=50$, $U=0.5$ eV \\
        \hline
        KIVC & $-14.171$ & $-3.579$ \\
        % \hline
        QAH & $-12.808$ & $-3.295$ \\
        % \hline
        OP & $-11.994$ & $-3.252$ \\
        % \hline
        SP & $-11.483$ & $-2.823$ \\
        % \hline
        NSM & $-9.835$ & $-2.812$ \\
        % \hline
        VPL & $-9.075$ & $-2.526$ \\
        % \hline
        TIVC & $-7.872$ & $-2.407$ \\
        \hline \hline
    \end{tabular}
    \caption{Condensation energy of the self-consistent states at charge neutrality relative to the energy of the symmetry-preserving normal state.}
    \label{tab1}
\end{table}

\section{Symmetry and correlated states}
% \paragraph*{\color{blue}{Symmetry and correlated states.}}

Here, we comment on the consequences of our renormalization procedure on the symmetry in the $n_B=4$ system. For a detailed analysis of the symmetries of MATBG, including larger $n_B$, we refer the reader to Appendix \ref{appd} as well as to Refs. \cite{bernevig21,Bultinck20,song22}.

The most important outcome is the widening of the flat bands by the large component ${h}_{x0}(\kk) \sigma_x + {h}_{yz}(\kk) \sigma_y \tau_z$. The generator $\mathcal{G}_{\text{flat}} = \sum_{\kk} c^\dagger_{\kk } \sigma_y \tau_x  c_{\kk }$ is then broken and the system diverges from the $U(4)_{\text{flat}}$ symmetry of the 'nonchiral-flat' limit. On the other hand, $\mathcal{G}_{\text{chiral}} = \sum_{\kk} c^\dagger_{\kk } \sigma_x \tau_y  c_{\kk }$ commutes with this dominant term thus the 'chiral-nonflat' $U(4)_{\text{chiral}}$ symmetry is approximately preserved.

With respect to the interaction, the breaking of $\mathcal{G}_{\text{chiral}}$ and $\mathcal{G}_{\text{flat}}$ can be assessed by the eigenvalues and singular values of $\overline{\sigma}_z$ and $C_{2z}\overline{P}$, respectively, displayed in Fig. \ref{fig2}(b); values less than one reveal the non-symmetry of the Hilbert space. These quantities increase in the renormalized state, thus enhancing the symmetry of ${\mathcal{H}}_{\text{int}}$.

% We draw attention to the 'nonchiral-flat' and 'chiral-nonflat' generators of the $U(4) \times U(4)$ symmetry,
% \begin{align}
%     \mathcal{G}_{\text{flat}} &= \sum_{\kk} c^\dagger_{\kk } \sigma_y \tau_x  c_{\kk }, \nonumber \\
%     \mathcal{G}_{\text{chiral}} &= \sum_{\kk} c^\dagger_{\kk } \sigma_x \tau_y  c_{\kk }.
% \end{align}

% The breaking of $\mathcal{G}_{\text{flat}}$ lowers $U(4)\times U(4)$ to the $U(4)_{\text{chiral}}$ subgroup. On the level of $\overline{\mathcal{H}}_{\text{int}}$, it can be assessed by the singular values of $C_{2z}\overline{P}$, values different from unity evidence the non-symmetry of the Hilbert space and consequently of the form factors (Eq. (\ref{formfactors})). Most importantly, the symmetry is broken by the large anticommuting term $\overline{h}_{x0}(\kk) \sigma_x + \overline{h}_{yz}(\kk) \sigma_y \tau_z$ of $\overline{h}$

% Alternatively, the (different from one) sublattice polarization of the narrow bands indicates breaking of $\mathcal{G}_{\text{chiral}}$; the $U(4) \times U(4)$ symmetry is then lowered to $U(4)_{\text{flat}}$.

% Notice that $|\text{KIVC} \rangle =  \text{exp}(i\pi/4 \ \mathcal{G}_{\text{chiral}})|\text{OP} \rangle$, $|\text{VP} \rangle =  \text{exp}(i\pi/4\ \mathcal{G}_{\text{chiral}})|\text{TIVC} \rangle$, $|\text{KIVC} \rangle =  \text{exp}(i\pi/4 \ \mathcal{G}_{\text{flat}})|\text{VP} \rangle$ and $|\text{TIVC} \rangle =  \text{exp}(i\pi/4 \ \mathcal{G}_{\text{flat}})|\text{OP} \rangle$. In the $U(4) \times U(4)$ limit all four are degenerate ground states \cite{lian21,Bultinck20,Hofmann22}.

In order to investigate the interacting phases, we have performed Hartree-Fock calculations in the flat-band theory. The self-consistent states can be described by the one-particle density matrix, $
    P(\kk)_{\alpha \beta} = \langle  c^\dagger_{\kk \alpha}  c_{\kk \beta} \rangle = \frac{1}{2}\Big( \mathbb{1}_{\alpha \beta} + Q(\kk)_{\alpha \beta} \Big)$.
% with the properties $P(\kk)^2 = P(\kk) = P(\kk)^\dagger$ and $\text{tr}(P(\kk)) =$ the number of occupied electrons at $\kk$. 
% The indices $\alpha, \beta = (\tau \sigma s)$ denote the valley-sublattice-spin flavor and $\langle ...  \rangle$ the expectation value. 

% For spin-singlets and 4 occupied electrons at each $\kk$, $Q(\kk)$ can be decomposed into products of $\sigma$ and $\tau$ matrices with real coefficients, $Q(\kk) = \sum_{ij} \langle\sigma_i \tau_j (\kk) \rangle\sigma_i\tau_j$. 

We study several symmetry-breaking phases at charge neutrality: the Kramers intervalley coherent state (KIVC, $Q(\kk)  = \sigma_y \tau_y$), orbital polarized (OP, $Q(\kk)  = \sigma_z$, also called valley Hall), valley polarized (VP, $Q(\kk) = \tau_z$), time-reversal invariant intervalley coherent (TIVC, $Q(\kk)  = \sigma_x \tau_x$) \cite{lian21,Bultinck20,Hofmann22}, Quantum Anomalous Hall (QAH, $Q(\kk) = \sigma_z \tau_z$) \cite{soejima20,Hofmann22,wang23,Bultinck20,lian21}, nematic semimetal (NSM, $Q(\kk)= \cos(\pi/3) \sigma_x - \sin(\pi/3) \sigma_y\tau_z$) \cite{soejima20,parker21,liu21,wang23} and the spin polarized state (SP, $Q(\kk) = s_z$, the spin-$z$ matrix). The states above serve as initial seeds in the self-consistency loop; for VP we enforced $C_{2z}\mathcal{T}$ and valley conservation symmetries, and for TIVC and NSM, $C_{2z}$ and $\mathcal{T}$. The energies of the converged states are reported in Table \ref{tab1}. 
 
\begin{figure}[t!]
     \centering
     \includegraphics[width=1\linewidth]{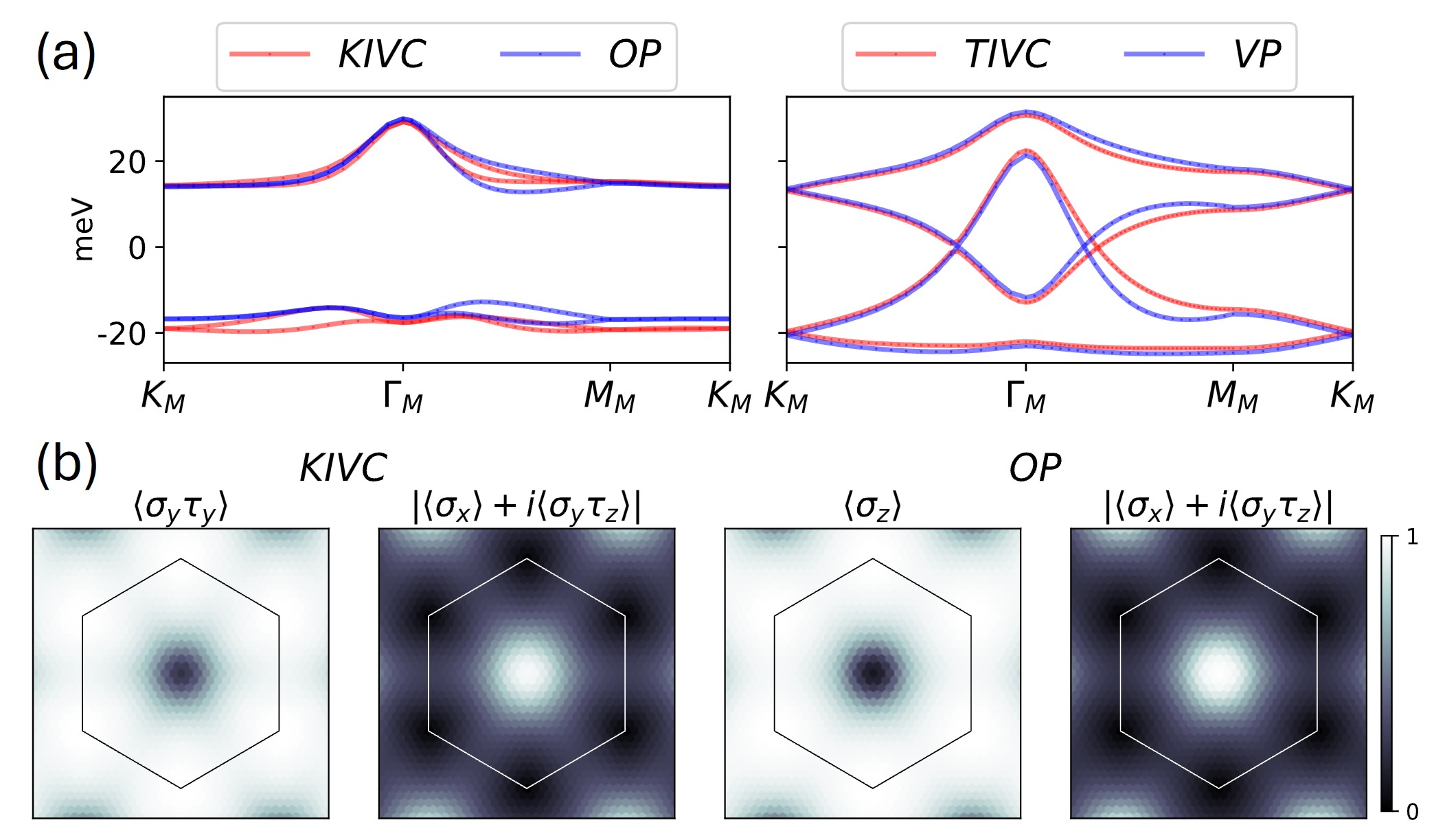}
     \caption{(a) Band structure of the Kramers intervalley coherent (KIVC), orbital polarized (OP), time-reversal invariant intervalley coherent (TIVC) and valley polarized (VP) states. The approximate chiral symmetry relating KIVC and OP, and TIVC and VP is manifest. (b) Order parameters, defined by $Q(\kk) = \sum_{ij} \langle\sigma_i \tau_j (\kk) \rangle\sigma_i\tau_j$, of the KIVC and OP states. The quantities $\langle \sigma_y\tau_y (\kk) \rangle$ and $\langle \sigma_z(\kk) \rangle$ are related by a chiral rotation, and the quantities $\langle \sigma_x (\kk) \rangle$ and $\langle \sigma_y\tau_z (\kk) \rangle$ are trivial under chiral rotations and correspond to filling the valence bands of $\mathcal{H}_0$. The coupling strength is $\epsilon_r=10$, $U=4$ eV.} 
     \label{fig4}
\end{figure}

% In the 'chiral-flat' $U(4) \times U(4)$ limit, KIVC, OP, VP and TIVC are degenerate \cite{lian21,Bultinck20,Hofmann22}. 
Invariably in all phases, the kinetic energy penalty induces $Q(\kk) \sim \sigma_x, \sigma_y\tau_z$ around the $\Gamma_M$ point, coming from the lower bands of $\mathcal{H}_0$. This is apparent in Fig. \ref{fig4}(b) where we plot the $Q(\kk)$ matrix of KIVC and OP. Concurrently, the wide dispersion enforces TIVC, VP and SP to be gapless, while KIVC, OP and QAH are insulating, see Fig. \ref{fig4}(a). These (semi)metallic states are destabilized by $\mathcal{H}_0$ as compared to the insulators and the NSM. Moreover, the approximate chiral symmetry, $|\text{KIVC} \rangle \approx  \text{exp}(i\pi/4 \ \mathcal{G}_{\text{chiral}})|\text{OP} \rangle$, $|\text{VP} \rangle \approx  \text{exp}(i\pi/4\ \mathcal{G}_{\text{chiral}})|\text{TIVC} \rangle$, is manifest in Fig. \ref{fig4}(a) and (b).
 
% If the $Q$ matrix at the BZ boundary commutes with $\sigma_x$ and $\sigma_y\tau_z$, then by the property $P(\kk) = P(\kk)^2$ a smooth evolution form valence-band to $Q$ order is forbidden; the transition has to be sharp, i.e. through a band crossing or avoided crossing. Heuristically, phases with $Q$ matrix anticommuting with $\sigma_x,\sigma_y\tau_z$ are insulating like KIVC and OP, and with $Q$ commuting are (semi-)metallic like (TIVC) VP and have higher energies.    

% In Table \ref{tabmain1} we report the energies of the Hartree-Fock fixed points relative to the energy of the symmetric state. Besides KIVC, OP, VP and TIVC, we obtained the self-consistent Quantum Anomalous Hall (QAH) state, with $Q = \sigma_z \tau_z$ \cite{soejima20,Hofmann22,wang23,Bultinck20,lian21}, the $C_{3z}$-breaking nematic semimetal (NSM), with $Q= \cos(\pi/3) \sigma_x - \sin(\pi/3) \sigma_y\tau_z$ \cite{soejima20,parker21,liu21,wang23,Bultinck20} and the spin polarized state (SP). 

% \paragraph*{\color{blue}{Discussion.}}
\section{Discussion}

% \begin{figure}[t]
%      \centering
%      \includegraphics[width=.99\linewidth]{fig4.jpg}
%      \caption{aa} 
%      \label{fig4}
% \end{figure}

In this work, we performed large-scale (11908 atoms per unit cell) computations in an atomistic model of MATBG and obtained a mean-field symmetric state at charge neutrality. In this state, the dispersion and the wave functions are renormalized such that the electrons are decoupled at the one-loop (Hartree-Fock) level. We argue that effective models with a reduced bandwidth reproducing this symmetric phase take account of the many-body effects coming from the removed degrees of freedom, as opposed to models obtained from a naive truncation of the bare Hilbert space.

We have constructed effective theories using a many-body projection onto the lowest $4$ and $20$ bands per spin of the symmetric state. The $4$-band model corresponds to the flat band projection, and the $20$-band model describes an energy window of about $\pm 250$ meV. 

In these theories the flat bands widen drastically, with a bandwidth comparable to the interaction scale, demonstrating the relevance of the renormalizations induced by integrating out the remote bands. Furthermore, the sublattice polarization of the flat-band subspace increases, showing that the renormalized system approaches the chiral limit of MATBG \cite{tarnopolski19,ledwith20,vafek20}.

The nonflat bands are directly linked to the breaking of the $U(4)_\text{flat}$ symmetry. Instead, $U(4)_\text{chiral}$ prevails as an approximate symmetry of the system, broken at a scale of some tens of meV$/\epsilon_r$, as manifested in the correlated states at charge neutrality.

In a previous study \cite{vafek20}, the authors studied the BM model with an energy cutoff of order of the graphene bandwidth, and integrated out the high-energy bands via a Renormalization Group procedure. However, the BM Hamiltonian is only accurate within a certain energy range, making the procedure uncontrolled above a threshold. In contrast, our scheme is self-consistent rather than perturbative, and more importantly, our tight-binding model accurately describes the wavefunctions within the full bandwidth of the $\pi$ orbitals. This explains the discrepancies in the final outcomes, for instance, the band widening in \cite{vafek20} is relatively weak.

On another note, our ordering of the Coulomb interaction (Eq. \ref{hamiltonian}) is consistent with the 'average' subtraction scheme in MATBG literature \cite{bernevig21,vafek20}. %,kang21,xie23}
Another widely used protocol is the 'graphene' subtraction scheme \cite{Bultinck20,xie20}, where ${H}_{\text{int}}$ is normal-ordered \cite{Giuliani_Vignale_2005} with respect to the ground state of two decoupled layers at neutrality.

It has been assumed that due to the subtraction, the normal state is close to the bare ground state in the graphene scheme \cite{Bultinck20,soejima20,parker21,Hofmann22,wang23,cea20}. Contrary to this expectation, our calculations with graphene subtraction, found in Appendix \ref{appe}, show significant band widening also; the flat bands reach a bandwidth of $43.8$ meV for $\epsilon_r=10$. Moreover, in this scheme we have identified a topological phase transition on the flat bands for some value of $\epsilon_r$ between $50$ and $10$.

% As a consequence of band renormalization, there is a strong tendency to populate the valence bands around $\Gamma_M$ in the interacting states. This fact is directly related to the breaking of the $U(4)_\text{flat}$ symmetry by the dispersion. Instead, we find that the chiral group $U(4)_\text{chiral}$ is an approximate symmetry of the system, broken at a scale of some tens of meV$/\epsilon_r$.

% To further discuss the implications of our findings in the interacting ground states, let focus on filling $\nu = -2$ electrons per unit cell, i.e. half filling of the valence bands. A previous STM study \cite{Nuckolls2023} identified a Kekule pattern in the local density of states of the $\nu=-2$ correlated insulator \cite{Cao2018_2,Zondiner2020,Saito2020,Saito2021,Cao2021,Lu2019,Stepanov2020}, pointing to time-reversal invariant intervalley coherence stabilized by the electron-phonon coupling \cite{wang24,kwan2023electronphononcouplingcompetingkekule,shi2024moireopticalphononsdancing}.

% The results presented here promise to have extensive consequences in our understanding of MATBG. We already mentioned the pattern of explicit symmetry breaking, $U(4)\times U(4) \to U(4)_{\text{chiral}}$ instead of $U(4)\times U(4) \to U(4)_{\text{flat}}$ which is conventionally assumed \cite{song22,hu23,hu23_2,wang2024electronphononcouplingtopological}. In close connection to this, the VP and TIVC states are 'dynamically enforced' to be gapless in contrast to most theory results \cite{song22,shi2024moireopticalphononsdancing,wang2024electronphononcouplingtopological}.
To conclude, let us mention that, in our model at $\theta=1.05^\circ$, the conduction (valence) states of the flat bands are mainly composed of conduction (valence) states of the parent graphene layers \cite{escudero24,yu23}; the Hartree-Fock potentials, hence, drive the graphene conduction-like sates upward in energy while the valence states are lowered. This suggests that the magic angle, as defined by the angle at which there is a band inversion at $\Gamma_M$ \cite{tarnopolski19,yu23,carr19,escudero24} (in our tight-binding model this occurs at $\theta=1.02^\circ$), gets renormalized towards a lower value \footnote{M. S\'anchez S\'anchez, T. Stauber and J. Gonz\'alez, in preparation. The magic angle is found to be between $0.9^\circ$ and $1.0^\circ$ for $\epsilon_r=10$, $U=4$ eV.}.

% In another regard, let us consider the strain-stabilized, moir\'e translations-breaking, inconmensurate Kekul\'e spiral (IKS) phase \cite{kwan21,wang23,wagner22,Nuckolls2023}. The IKS momentum, $\boldsymbol{q}_{\text{IKS}}$, is determined by a nesting condition in the band structure of the nematic semimetal (NSM), and it can be measured in STM experiments \cite{Kim2023}. The NSM can be significantly modified by the band renormalization, for instance the high energy cost if the Dirac nodes lie close to $\Gamma_M$ from $\mathcal{H}_0$ competes with the exchange gain if the Dirac cones lie at the BZ center \cite{liu21}. Thus, a comparison between the theoretically predicted and the experimentally determined $\boldsymbol{q}_{\text{IKS}}$ provides a direct test of our findings.

\section*{Acknowledgments}
We thank Patrick Ledwith and Eslam Khalaf for fruitful discussions. This work was supported by grant PID2020–113164 GBI00 funded by MCIN/AEI/10.13039/501100011033, PID2023-146461NB-I00 and PRE2021-097070 funded by Ministerio de Ciencia, Innovación y Universidades, and the CSIC Research Platform on Quantum Technologies PTI-001. The access to computational resources of CESGA (Centro de Supercomputaci\'on
de Galicia) is also gratefully acknowledged.

% \nocite{*}
\bibliography{biblio}

%%%%%%%%%%%%%%%%%%%%%%%%%%%%%%%%%%%%%%%%%%%%%%%%%%%%%%%%%

\clearpage

\appendix

% \title{Supplementary Material for "Nonflat bands and chiral symmetry in magic-angle twisted bilayer graphene"}

% \author{Miguel Sánchez Sánchez}
% \email{miguel.sanchez@csic.es}
% \affiliation{Insituto de Ciencia de Materiales de Madrid ICMM-CSIC. Madrid (Spain).}
% \author{José González}
% \affiliation{Insituto de Estructura de la Materia IEM-CSIC. Madrid (Spain)}
% \author{Tobias Stauber}
% \email{tobias.stauber@csic.es}
% \affiliation{Insituto de Ciencia de Materiales de Madrid ICMM-CSIC. Madrid (Spain).}

% %\keywords{Suggested keywords}%Use showkeys class option if keyword
%                               %display desired
% \maketitle

% \setcounter{equation}{0}
% \renewcommand\theequation{S\arabic{equation}}

\setcounter{figure}{0}
\renewcommand\thefigure{\thesection\arabic{figure}}    

\setcounter{table}{0}
\renewcommand\thetable{\thesection\arabic{table}}

% \title{Supplementary Material for "Nonflat bands and chiral symmetry in magic-angle twisted bilayer graphene"}
% \maketitle

\onecolumngrid

% \tableofcontents

\clearpage

\section{Lattice geometry and tight-binding model}\label{appa}

In a graphene monolayer the lattice vectors are $\boldsymbol{a}_1 = a_0(1/2,\sqrt{3}/2)$ and $\boldsymbol{a}_2 = a_0(-1/2,\sqrt{3}/2)$, with $a_0 = 2.46$ $\text{\AA}$ the lattice constant. The atoms on the latice positions set up the $A$ sublattice, and the atoms displaced from the $A$ atoms by $(\boldsymbol{a}_1 + \boldsymbol{a_2})/3$  set up the $B$ sublattice. The dual lattice vectors are then $\boldsymbol{G}_1 = 4\pi/\sqrt{3}a_0 (\sqrt{3}/2,1/2)$ and $\boldsymbol{G}_2 = 4\pi/\sqrt{3}a_0 (-\sqrt{3}/2,1/2)$. The band structure displays the characteristic Dirac cones around the two $K$ points, $K = (\boldsymbol{G_2} - \boldsymbol{G_1})/3$ and $K' = -K$. Our convention differs from other works in which $K = (\boldsymbol{G_1} - \boldsymbol{G_2})/3$ by a $180^\circ$ rotation, or equivalently by an interchange of the $A$ and $B$ sublattice labels.

We set up the twisted bilayer geometry starting from a graphene bilayer with the two layers stacked on top of each other, at vertical positions $z=\pm d_0/2 = \pm 1.6755$ $\text{\AA}$. The top layer gets rotated by an angle $\theta/2$ and the bottom layer by $-\theta/2$, with the center of rotation being the center of one of the carbon hexagons. For a generic twist angle $\theta$, the resulting structure is not strictly periodic, only for particular angles there is a commensurate crystalline structure. A subset of such commensurate angles can be parameterized as $\cos(\theta) = 1 - 1/(6n^2 +6n +2)$ with integer $n$ \cite{dossantos12}. The unit vectors of the resulting superlattice are $\boldsymbol{L}_1 = L_M(\sqrt{3}/2,1/2)$ and $\boldsymbol{L}_2 = L_M(-\sqrt{3}/2,1/2)$ with $L_M = a_0\sqrt{3n^2+3 n+ 1}$ the moiré lattice constant. The reciprocal lattice in turn has unit vectors $\boldsymbol{g}_1 = 4\pi/\sqrt{3}L_M(1/2,\sqrt{3}/2)$ and $\boldsymbol{g}_2 = 4\pi/\sqrt{3}L_M(-1/2,\sqrt{3}/2)$.

The magic-angle of twisted bilayer graphene sits between $1.0^\circ$ and $1.1^\circ$ in different experiments and theoretical models. In this work we chose $\theta = 1.05 ^\circ$ ($n=31$). The moiré lattice constant is of $13.4$ nm and the number of atoms in the unit cell is $11908$. We have included the effects of in-plane lattice relaxation, i.e. symmetry-preserving intrinsic strain, using the model of Ref. \cite{nam17}. 

In the bilayer, the crystallographic symmetry of $6$-fold rotations around the twist axis (the $z$ axis through one of the hexagons), denoted $C_{6z}$, is preserved. Yet the mirror symmetry about the $y$ axis of each individual layer is lost in the twisted structure, as it relates a lattice with twist angle $\theta$ to one with angle $-\theta$. Instead, there is a related symmetry consisting of $180^\circ$ rotations around the $y$ axis at $z=0$, denoted $C_{2y}$. This operation acts as a mirror symmetry about the $y$ axis plus an interchange of the layers. The group elements $C_{6z}^2$, $C_{6z}^3$ and $C_{2z}C_{2y}$ performing $3$-fold $z$-rotations, $2$-fold $z$-rotations and $180^\circ$ rotations around the $x$ axis are denoted $C_{3z}$, $C_{2z}$ and $C_{2x}$, respectively. Furthermore, the spin-orbit coupling in carbon is negligible, so there is an additional spinless time-reversal symmetry, $\mathcal{T}$, acting as complex conjugation.

We set up a microscopic tight-binding model for the carbon $\pi$ orbitals of MATBG. We employ the Slater-Koster \cite{slater54} hopping integral given in Refs. \cite{laissandiere10,moon12}. The hopping function depends on the distance between atoms and the angle of the radius vector with the $z$ axis,
\begin{equation}
\begin{gathered}
    % t(\boldsymbol{r}) = - V_{pp\pi}(r) \Bigg(1 - \bigg(\frac{\boldsymbol{r}\cdot \boldsymbol{\hat{z}}}{ r}\bigg)^2\Bigg) + V_{pp\sigma}(r) \bigg(\frac{\boldsymbol{r}\cdot \boldsymbol{\hat{z}}}{ r}\bigg)^2 \nonumber \\
    % V_{pp\pi}(r) = V_{pp \pi}^0 e^{-(r - a_0)/r_0} \quad \quad \quad V_{pp\sigma}(r) = V_{pp \sigma}^0 e^{-(r - d_0)/r_0} \nonumber \\
    % V^0_{ppp\pi} = 2.7 \text{eV} \quad \quad \quad \quad V^0_{pp\sigma} = 0.48 \text{eV}
    t(\boldsymbol{r}) = - V_{pp\pi}(r) \big(1 - \cos^2(\varphi_{\boldsymbol{r}} )\big) + V_{pp\sigma}(r) \cos^2(\varphi_{\boldsymbol{r}}), \\
    V_{pp\pi}(r) = V_{pp \pi}^0 \ \text{exp}\big(-(r - a_{cc})/r_0\big), \quad \quad \quad V_{pp\sigma}(r) = V_{pp \sigma}^0 \ \text{exp}\big(-(r - d_0)/r_0\big),
\end{gathered}
\end{equation}
where $\boldsymbol{r}=(x,y,z)$, $r=\sqrt{x^2+y^2+z^2}$ and $\cos(\varphi_{\boldsymbol{r}}) = z/r$. The constants are $a_{cc} = a_0/\sqrt{3}$ (the carbon-carbon distance), $V^0_{ppp\pi} = 2.7$ eV,  $V^0_{pp\sigma} = 0.48$ eV and $r_0 = 0.0453 $ nm. 

\begin{figure*}[b]
    \centering
    \includegraphics[width=.9\linewidth]{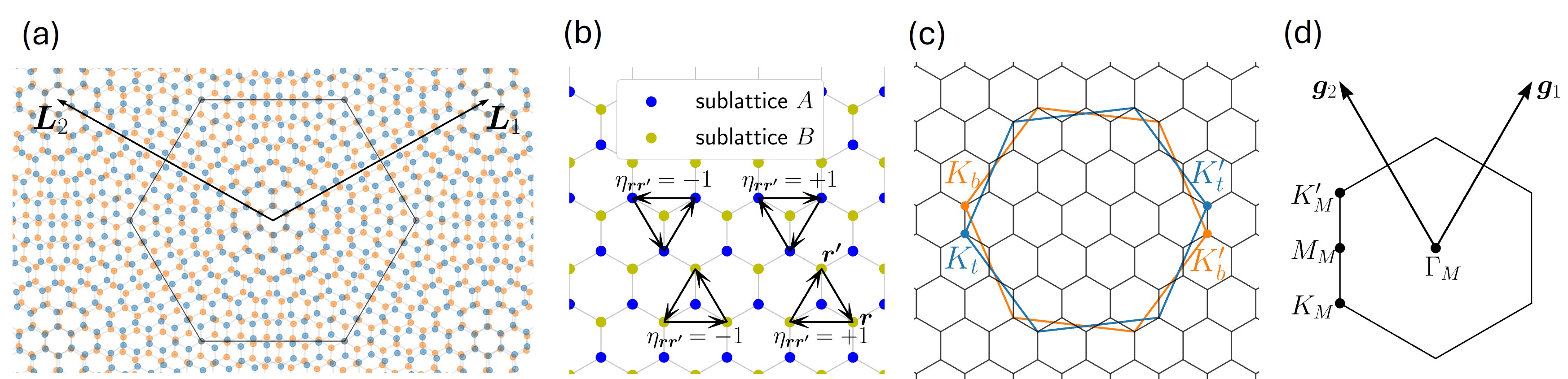}
    \caption{(a) Top view of MATBG with a twist angle of $6.01^\circ$ ($n=5$). The lattice vectors $\boldsymbol{L}_1$, $\boldsymbol{L}_2$ and the moiré Wigner-Seitz cell are marked. (b) Triangular loops used for computing the valley charge. (c) Brillouin zones of the rotated top (blue) and bottom (orange) monolayers, and the hexagonal Brillouin zones of the coupled layers (black), drawn in an extended scheme. The $K(K')$ points of the original graphenes are depicted. (d) Moiré Brillouin zone of MATBG, with the high-symmetry $\Gamma_M$, $M_M$, $K_M$, $K'_M$ points and the reciprocal lattice vectors $\boldsymbol{g}_1$, $\boldsymbol{g}_2$ labeled.} 
    \label{fig5}
\end{figure*}

\clearpage
\section{Mean-field decoupling and many-body projection}\label{appb}

Let us recall the full interacting tight-binding Hamiltonian,
\begin{align}
\begin{split}
    & {H} =  H_{\text{TB}} + {H}_\text{int}  \\ 
    & H_{\text{TB}} = \sum_{\boldsymbol{r}\boldsymbol{r'}s} t(\boldsymbol{r}-\boldsymbol{r'}) c^\dagger_{\boldsymbol{r}s}c_{\boldsymbol{r'}s}   \\  
    & {H}_\text{int} = \frac{1}{2}\sum_{\boldsymbol{r}\neq \boldsymbol{r'},ss'} V(\boldsymbol{r}- \boldsymbol{r'}) \bigg(c^\dagger_{\boldsymbol{r}s}  c_{\boldsymbol{r}s} - \frac{1}{2}\bigg)  \bigg(c^\dagger_{\boldsymbol{r'}s'}  c_{\boldsymbol{r'}s'} - \frac{1}{2}\bigg) + \frac{1}{2}\sum_{\boldsymbol{r}s} U  \bigg(c^\dagger_{\boldsymbol{r}s}  c_{\boldsymbol{r}s} - \frac{1}{2}\bigg)  \bigg(c^\dagger_{\boldsymbol{r}\Bar{s}}  c_{\boldsymbol{r}\Bar{s}} - \frac{1}{2}\bigg),
    \end{split}
    \label{hamiltoniansm}
\end{align}
with $V(\boldsymbol{r}-\boldsymbol{r'}) = \frac{14.4 \text{eV}\times\text{\AA}}{\epsilon_r}\sum_{n=-\infty}^\infty \frac{(-1)^n}{||\boldsymbol{r} - \boldsymbol{r'} + 100 \text{\AA} \times n \boldsymbol{\hat{z}} ||}$ and $\Bar{s}$ the spin projection opposite to $s$. We solve $H$ using the standard mean-field decoupling \cite{Giuliani_Vignale_2005},
\begin{align}
    {H} \approx {H}_{\text{MF}} =& \sum_{\boldsymbol{r}\boldsymbol{r'}s} t(\boldsymbol{r} - \boldsymbol{r'}) c^\dagger_{\boldsymbol{r}s}c_{\boldsymbol{r'}s} + \sum_{\boldsymbol{r}\neq \boldsymbol{r'},ss'} V(\boldsymbol{r}- \boldsymbol{r'}) \bigg(\langle c^\dagger_{\boldsymbol{r}s'} c_{\boldsymbol{r}s'} \rangle_0  - \frac{1}{2} \bigg) c^\dagger_{\boldsymbol{r}s} c_{\boldsymbol{r}s} \nonumber \\ 
    &-\sum_{\boldsymbol{r}\neq \boldsymbol{r'},s} V(\boldsymbol{r}- \boldsymbol{r})  \langle c^\dagger_{\boldsymbol{r'}s} c_{\boldsymbol{r}s} \rangle_0 c^\dagger_{\boldsymbol{r}s} c_{\boldsymbol{r'}s} + \sum_{\boldsymbol{r} s} U \bigg( \langle c^\dagger_{\boldsymbol{r}\Bar{s}} c_{\boldsymbol{r}\Bar{s}} \rangle_0  - \frac{1}{2} \bigg) c^\dagger_{\boldsymbol{r}s} c_{\boldsymbol{r'}s} + \text{constant},
    \label{H_MF}
\end{align}
were $\langle ... \rangle_0$ denotes the expectation value in a state of reference. We look for a self-consistent solution where the reference state corresponds to the ground state of ${H}_{\text{MF}}$. We assume a spin-symmetric state, hence  $\langle c^\dagger_{\boldsymbol{r}s} c_{\boldsymbol{r'}\Bar{s}} \rangle_0 = 0$ and the spins are decoupled in the Fock channel (third term). 

In practice, we assume periodic boundary conditions and preserved moiré translations, and work in the basis of Bloch waves,
\begin{align}
c^\dagger_{\boldsymbol{k}\boldsymbol{\delta} s} = \frac{1}{\sqrt{N_c}}\sum_{\boldsymbol{R}_\ell} e^{i\kk \cdot (\boldsymbol{\delta} + \boldsymbol{R}_\ell)} c^\dagger_{\boldsymbol{\delta}+ \boldsymbol{R}_\ell, s},    
\end{align}
with $N_c$ the number of unit cells and $\kk$ is the Bloch momentum inside the rhombic Brillouin zone, $ \kk = (n_1/\sqrt{N_c}) \boldsymbol{G_1} + (n_2/\sqrt{N_c}) \boldsymbol{G_2}$ for integers $n_1,n_2=0,...,\sqrt{N_c}-1$. The index $\boldsymbol{\delta}$ denotes an atomic position inside the Wigner-Seitz cell, and $\boldsymbol{R}_\ell$ denotes the lattice vectors. Different Bloch momenta are coupled in ${H}$, but by translation symmetry they become decoupled in the mean-field Hamiltonian ${H}_\text{MF}$. The mean-field basis $c^\dagger_{\boldsymbol{k}n s}$ diagonalizes ${H}_{\text{MF}}$,
\begin{align}
\begin{split}
    {H}_{\text{MF}} =& \sum_{\boldsymbol{k}ns} \varepsilon^{\text{MF}}_n(\kk) c^\dagger_{\boldsymbol{k}n s} c_{\boldsymbol{k}ns} + \text{constant}, \\  
    c^\dagger_{\boldsymbol{k}n s} =& \sum_{\boldsymbol{\delta}}\langle \kk \boldsymbol{\delta} | \boldsymbol{k} n \rangle  c^\dagger_{\boldsymbol{k}\boldsymbol{\delta}s}
    % = \frac{1}{\sqrt{N_c}} \sum_{\boldsymbol{r}} e^{i \boldsymbol{k} \cdot \boldsymbol{r}} \psi_{\boldsymbol{k}n\boldsymbol{r}} c^\dagger_{\boldsymbol{r}s}
    = \sum_{\boldsymbol{r}} \langle \boldsymbol{r} | \boldsymbol{k} n \rangle c^\dagger_{\boldsymbol{r}s},
\end{split}
\end{align}
where $\langle \boldsymbol{k} \boldsymbol{\delta} | \boldsymbol{k} n \rangle $ and $ \langle \boldsymbol{r} | \boldsymbol{k} n \rangle$ are the wave function amplitudes (independent of spin) of mode $c^\dagger_{\boldsymbol{k}ns}$ in the Bloch and real-space basis, respectively. The conduction bands have band index $n=1,2,...$ with increasing $n$ corresponding to increasing energy, and $n=-1,-2,...$ denotes the valence bands with increasing $|n|$ corresponding to decreasing energy. In the self-consistent algorithm we used the ground state of the bare system ($H_\text{TB}$) as initial seed and imposed the crystallographic symmetries and spinless time-reversal symmetry at each iteration. We used a $6\times 6$ discretization of the Brillouin zone ($N_c=36$); results for finer momentum grids are obtained from $H_\text{MF}$ (Eq. \ref{H_MF}) with the self-consistent Fock matrix, $\langle c^\dagger_{\boldsymbol{r}s} c_{\boldsymbol{r'}s'} \rangle_0$, of the $6\times6$ system. The ODA algorithm \cite{Kudin2002} is used to accelerate convergence.

With the renormalized symmetric state at hand, we wish to construct an effective model that accounts for the relevant low-energy excitations accurately. In order to achieve this, we assume that the mean-field bands with $n  < -n_B/2$ are fully occupied and with $n > n_B/2$ fully empty, and we allow for fluctuations in the central $n_B$ bands. We restrict to such states via the the many-body projector
\begin{align}
\begin{split}
        \mathcal{P} =& \sum_{I} | \Psi_I \rangle \big\langle \Psi_I |, \\  
    |\Psi_I \rangle =& \prod_{\kk n s \in I} c^\dagger_{\boldsymbol{k} n s} \prod_{\boldsymbol{q} s', m < {-n_B}/{2}}c^\dagger_{\boldsymbol{q} m s'} |0\rangle,
\end{split}    
\label{projwf}
\end{align}
where $I$ denotes the set of indices of one particular element of a complete basis of the subspace of many-body excitations of the central $n_B$ bands, and the sum over $I$ runs over all $2^{N_C \times n_B\times 2}$ basis elements. In other words, $\mathcal{P}$ projects to all $n$-body ($n=1,2,...,N_c\times n_B \times 2$) excitations on the central $n_B$ bands.

Consider a generic two-body Hamiltonian $V = \frac{1}{2}\sum_{abcd}V_{abcd}c^\dagger_a c^\dagger_b c_c c_d$. (in our case, $V=\frac{1}{2}\sum_{\boldsymbol{r}\neq \boldsymbol{r'},ss'} V(\boldsymbol{r}- \boldsymbol{r'})c^\dagger_{\boldsymbol{r}s} c^\dagger_{\boldsymbol{r'}s'}  c_{\boldsymbol{r'}s'} c_{\boldsymbol{r}s} + \frac{1}{2} \sum_{\boldsymbol{r}s} U c^\dagger_{\boldsymbol{r}s} c^\dagger_{\boldsymbol{r}\Bar{s}}c_{\boldsymbol{r}\Bar{s}} c_{\boldsymbol{r}s}$). We can deduce the form of $\mathcal{P} V \mathcal{P}$ by considering the action of $V$ on a generic two-particle state in the central bands. We will work in the mean-field basis and use $a,b,c,d$ for generic modes, $\alpha ,\beta,\gamma,\delta$ for modes within the central bands and $i,j$ for modes with $n<-n_B/2$. After some algebra, we get
\begin{align}
    \frac{1}{2} \sum_{abcd}V_{abcd}c^\dagger_a c^\dagger_b c_c c_d \ c_\alpha^\dagger  c_\beta^\dagger \prod_{j} c^\dagger_j \big|0\big\rangle =& \frac{1}{2} \sum_{\gamma\delta} (V_{\gamma\delta\beta\alpha} - V_{\gamma\delta\alpha\beta}) c_\gamma^\dagger c_\delta^\dagger \prod_{j} c^\dagger_j \big|0\big\rangle \nonumber \\
    &+ \sum_{\gamma i} (V_{\gamma i i \alpha} - V_{i \gamma i \alpha}) c_\gamma^\dagger c_\beta^\dagger\prod_j c^\dagger_j \big|0\big\rangle   - \sum_{\gamma i} (V_{\gamma i i \beta} - V_{i \gamma i \beta}) c_\gamma^\dagger c_\alpha^\dagger\prod_j c^\dagger_j \big|0\big\rangle \nonumber \\
    &+ \frac{1}{2}\sum_{ik} (V_{ikki} - V_{ikik}) c^\dagger_\alpha c^\dagger_\beta \prod_j c^\dagger_j \big|0\big\rangle + \text{others}, 
    \label{projV1}
\end{align}
where again we have restricted the final states to be of the form $ c_\alpha^\dagger  c_\beta^\dagger \prod_{j} c^\dagger_j |0\rangle$; different states are collectively denoted as 'others'. Hence, the many-body projected $V$ reads
\begin{align}
     \mathcal{P} V \mathcal{P} = \frac{1}{2} \sum_{\alpha\beta\gamma\delta} V_{\alpha\beta\gamma\delta}c^\dagger_\alpha c^\dagger_\beta c_\gamma c_\delta + \sum_{\alpha\beta} \sum_i (V_{\alpha i i \beta} - V_{i\alpha i \beta}) c^\dagger_\alpha c_\beta + \frac{1}{2} \sum_{ik} (V_{ikki} - V_{ikik})
     \label{projV2}
\end{align}
Using Eq. (\ref{projV1}) one can check that the expression in Eq. (\ref{projV2}) has the same action as $V$ if we restrict to initial and final states contained in $\mathcal{P}$. One now sees that, crucially, the occupied modes that have been integrated out induce new terms which are precisely the Hartree and Fock potentials acting on the central bands. The constant piece is the energy of the projected-out Fermi sea.

On the other hand, on the operator $O = \sum_{ab} O_{ab} c^\dagger_a c_b$ the projection truncates to the $\alpha$ modes and adds a constant,
\begin{align}
     \mathcal{P} O \mathcal{P} =\sum_{\alpha \beta}  O_{\alpha \beta} c^\dagger_\alpha c_\beta + \sum_{i} O_{ii}.
     \label{singlepartopproj}
\end{align}

We have just deduced that the effective model with Hamiltonian $\mathcal{P} {H} \mathcal{P}$ accounts for the many-body effects of the integrated valence bands, $n<-n_B/2$, via the one-particle terms in $\mathcal{P} {H}_{\text{int}} \mathcal{P}$. Now, remember that the dispersion of the symmetric state $\varepsilon^{\text{MF}}_n(\kk)$ already includes the mean-field potentials from the valence bands $n<0$, hence the projected Hamiltonian can be written in terms of the dispersion $\varepsilon^\text{MF}_n(\kk)$ plus a counter term from the valence bands with $0>n \geq -n_B/2$ to avoid double counting,
\begin{align}
    {\mathcal{H}} = \mathcal{P} {H} \mathcal{P} =& {\sum_{\kk n s }}' \varepsilon^\text{MF}_n(\boldsymbol{k}) c^\dagger_{\kk n s} c_{\kk n s} + \frac{1}{2}\sum_{\boldsymbol{r}\neq \boldsymbol{r'},ss'} V(\boldsymbol{r} - \boldsymbol{r'}) \ d^\dagger_{\boldsymbol{r}s} d^\dagger_{\boldsymbol{r'}s'} d_{\boldsymbol{r'}s'} d_{\boldsymbol{r}s}
    + \sum_{\boldsymbol{r}} U \ d^\dagger_{\boldsymbol{r}\uparrow} d^\dagger_{\boldsymbol{r}\downarrow} d_{\boldsymbol{r}\downarrow} d_{\boldsymbol{r}\uparrow} \nonumber \\
    &+ \sum_{\boldsymbol{r}\neq \boldsymbol{r'}, s} V(\boldsymbol{r}- \boldsymbol{r'})  \bigg(  \sideset{}{'}\sum_{\boldsymbol{q}, m<0} \langle \boldsymbol{r} | \boldsymbol{q} m \rangle  \langle \boldsymbol{q} m | \boldsymbol{r'}  \rangle \bigg)d^\dagger_{\boldsymbol{r}s} d_{\boldsymbol{r'}s} \nonumber \\ &-  2\sum_{\boldsymbol{r}\neq \boldsymbol{r'},s} V(\boldsymbol{r}- \boldsymbol{r'})\rho_{0\boldsymbol{r'}}\ d^\dagger_{\boldsymbol{r}s} d_{\boldsymbol{r}s} - \sum_{\boldsymbol{r}} U \rho_{0\boldsymbol{r}}\ d^\dagger_{\boldsymbol{r}s} d_{\boldsymbol{r}s} \nonumber \\ &+ \text{constant}, 
\end{align}
with 
\begin{gather}
      d^\dagger_{\boldsymbol{r}s} = {\sum_{\kk n}}' \langle \kk n | \boldsymbol{r} \rangle c^\dagger_{\kk n s},\quad \quad \quad      
      % \rho_{0\boldsymbol{r}}  = \sideset{}{'} \sum_{\kk,n<0} |\langle  \boldsymbol{r} |  \kk n  \rangle|^2.      
      \rho_{0\boldsymbol{r}}  = {\sum_{\kk,n<0}}' |\langle  \boldsymbol{r} |  \kk n  \rangle|^2.
\end{gather}
The primed summations indicate that $n$ is restricted to the central bands, $n=-n_B/2,...,n_B/2$. The Fock and Hartree counter-terms are written in the second and third line, respectively. The 'constant' is the zero-point energy of the frozen Fermi sea.

Now, the truncated position operators satisfy the relations $\{d^\dagger_{\boldsymbol{r}s}, d^\dagger_{\boldsymbol{r'}s'}\} = \{d_{\boldsymbol{r}s},d_{\boldsymbol{r'}s'}\} = 0$, $\{d^\dagger_{\boldsymbol{r}s}, d_{\boldsymbol{r'}s'}\} = \sum_{\kk n}' \langle \boldsymbol{r'} | \kk n \rangle \langle \kk n | \boldsymbol{r} \rangle \delta_{s s'}$. In order to write the interacting term in density-density form ($\sim d^\dagger_{\boldsymbol{r}s} d_{\boldsymbol{r}s} d^\dagger_{\boldsymbol{r'}s'}d_{\boldsymbol{r'}s'}$) we must take account of the anomalous anticommutator when changing the order of operators. In sum, the projected Hamiltonian reads
\begin{align}
\begin{split}
    &{\mathcal{H}} = {\mathcal{H}}_0 + {\mathcal{H}}_{\text{int}} + \text{constant},  \\
    &{\mathcal{H}}_0 = {\sum_{\kk n m  s}}' {h}(\boldsymbol{k})_{nm} c^\dagger_{\kk n s} c_{\kk m s},  \\ 
    &{\mathcal{H}}_{\text{int}} = \frac{1}{2}\sum_{\boldsymbol{r}\neq \boldsymbol{r'},ss'} V(\boldsymbol{r}- \boldsymbol{r'}) \delta\rho_{\boldsymbol{r}s} \delta\rho_{\boldsymbol{r'}s'} 
    + \sum_{\boldsymbol{r}} U \delta\rho_{\boldsymbol{r}\uparrow} \delta\rho_{\boldsymbol{r}\downarrow}.
    \end{split}
    \label{projhamiltoniansm}
\end{align}
The exchange components have been combined with $\varepsilon^\text{MF}_n(\kk)$ in ${\mathcal{H}}_0$, while the Hartree components are implicit in ${\mathcal{H}}_{\text{int}}$ through the background density $\rho_{0\boldsymbol{r}}$,
\begin{equation}
\begin{gathered}
    h(\boldsymbol{k})_{nm} = \varepsilon^{\text{MF}}_n(\kk) \delta_{nm}+ \frac{1}{2} \sum_{\boldsymbol{r}\neq \boldsymbol{r'}} V(\boldsymbol{r}- \boldsymbol{r'}) \langle \kk n | \boldsymbol{r} \rangle \langle \boldsymbol{r'} | \kk m \rangle \bigg( \sideset{\,}{'} \sum_{\boldsymbol{q},\ell<0} \langle \boldsymbol{r} |  \boldsymbol{q} \ell \rangle  \langle \boldsymbol{q} \ell | \boldsymbol{r'} \rangle - \sideset{}{'} \sum_{\boldsymbol{q},\ell>0}  \langle \boldsymbol{r} |  \boldsymbol{q} \ell \rangle  \langle \boldsymbol{q} \ell | \boldsymbol{r'} \rangle \bigg),  \\
     \delta\rho_{\boldsymbol{r}s} = \rho_{\boldsymbol{r}s} - \rho_{0\boldsymbol{r}}, \quad \quad \quad \quad \quad \rho_{\boldsymbol{r}s} = d^\dagger_{\boldsymbol{r}s} d_{\boldsymbol{r}s}.
\end{gathered}
\label{projhamiltoniansm2}
\end{equation}
This is the form of the effective models studied in the main text.

% \clearpage
\section{Vortex Chern basis of the flat bands}\label{appc}

In this section we describe a gauge-fixing procedure on the flat-band manifold. We follow the procedure of Ref. \cite{Bultinck20}, constructing the 'sublattice basis', and use the conventions of Ref. \cite{bernevig21}  for the definition and gauge fixing of the 'particle-hole' operator, see below. This 'Chern gauge' was originally defined for the bare flat bands, but we can equally impose it for the central bands of the mean-field state because both systems are adiabatically connected. We assume a spinless system; the physical system with spin includes two identical copies of the basis below.

Consider the microscopic valley ($\tilde{\tau}_z$) \cite{lopez20} and sublattice  ($\tilde{\sigma}_z$) operators,
\begin{align}
\begin{split}
    \tilde{\tau}_z &= \frac{i}{3\sqrt{3}}\Bigg( \sum_{\substack{\ll \boldsymbol{r},\boldsymbol{r'} \gg, s \\ \boldsymbol{r},\boldsymbol{r'} \in \text{sl} \ A}} \eta_{\boldsymbol{rr'}}c^\dagger_{\boldsymbol{r}s}c_{\boldsymbol{r'}s} - \sum_{\substack{\ll \boldsymbol{r},\boldsymbol{r'} \gg, s \\ \boldsymbol{r},\boldsymbol{r'} \in \text{sl} \ B}} \eta_{\boldsymbol{rr'}}c^\dagger_{\boldsymbol{r}s}c_{\boldsymbol{r'}s}\Bigg), \\
    \tilde{\sigma}_z &= \sum_{\boldsymbol{r} \in \text{sl}\ A,s} c^\dagger_{\boldsymbol{r}s} c_{\boldsymbol{r}s} - \sum_{\boldsymbol{r}\in \text{sl}\ B,s} c^\dagger_{\boldsymbol{r}s} c_{\boldsymbol{r}s},
    \label{vslop}
\end{split}
\end{align}
where $\ll \boldsymbol{r},\boldsymbol{r'} \gg$ denotes next-nearest neighbors, 'sl' is short for sublattice and $\eta_{\boldsymbol{rr'}} = \pm 1$ depending if the pair $\boldsymbol{r},\boldsymbol{r'}$ is part of a clockwise or counterclockwise triangle, see Fig. \ref{fig5}(b). Both operators conserve the Bloch momentum and $[\tilde{\tau}_z,\tilde{\sigma}_z]=0$. The valley operator is constructed so that graphene plane waves near the $K$($K'$) point have expectation values close to $1$($-1$).

Here and throughout, single-particle operators restricted to the flat bands will be indicated by a line, '$\overline{\phantom{o}}$' (Eq. (\ref{singlepartopproj}) would then read $\mathcal{P} O \mathcal{P} = \overline{O} + \sum_i O_{ii}$). We start by simultaneously diagonalizing the flat-band matrices at each $\kk$, $[\overline{\sigma}_z(\kk)]_{nm} = \langle \kk n | \tilde{\sigma}_z | \kk m \rangle$ and $[\overline{\tau}_z(\kk)]_{nm} = \langle \kk n | \tilde{\tau}_z | \kk m \rangle$. Valley $K$($K'$) states are identified as the states with positive (negative) eigenvalue under $\overline{\tau}_z$. 
We use $\tau_{i}$, $\sigma_i$ for the identity ($i=0$) and Pauli ($i=x,y,z$) matrices acting in the valley and sublattice polarized sectors, respectively. Valley $K(K')$ has eigenvalue $1(-1)$ under $\tau_z$, and we assign eigenvalue $1(-1)$ to sublattice $A(B)$ under $\sigma_z$. The resulting bands are topological with Chern numbers equal to $\sigma_z\tau_z$ \cite{Bultinck20,liu19,song21}.

The 'particle-hole' operator, $\tilde{P}$, is emergent in the continuum limit of MATBG \cite{song19,song21,kang23,herzogarbeitman2024heavyfermionsefficientrepresentation}. In practice, we consider the product $C_{2z}\tilde{P}$ rather than $\tilde{P}$ (understand the product of single-particle operators in second quantization as the second-quantized version of the product of the corresponding first-quantized operators). For ease of notation we will not distinguish between the microscopic, indicated with '$\tilde{\phantom{o}}$', and projected, indicated with '$\overline{\phantom{o}}$', versions of the crystallographic and time-reversal symmetries; their distinction must be understood within the context. Note moreover that the exact symmetries are block-diagonal in the band basis. For the definition of $C_{2z}\tilde{P}$, see Appendix \ref{appd}. 

The symmetry $C_{2z}\mathcal{T}$ and the particle-hole operator are local in Bloch momentum. We impose the 'Chern gauge' on the valley-sublattice basis, defined by the following form of $C_{2z}\mathcal{T}$ and $C_{2z}\overline{P}$,
\begin{align}
\begin{split}
    C_{2z}\mathcal{T} = \sigma_x \mathcal{K},    \\
    C_{2z}\overline{P} = \sigma_y \tau_x,
    \label{cherngauge}
\end{split}
\end{align}
with $\mathcal{K}$ denoting the antilinear complex conjugation operator. This gauge respects the relations $\{C_{2z}\overline{P},\tau_z\} = [C_{2z}\overline{P}, C_{2z}\mathcal{T}] = \{C_{2z}\overline{P}, \overline{\sigma}_z \} = [C_{2z}\mathcal{T}, \tau_z] =  \{ C_{2z}\mathcal{T},  \overline{\sigma}_z \} = 0$.

As a matter of fact, the flat Hilbert space is not invariant under the action of the particle-hole operator (in other words, $C_{2z}\tilde{P}$ does not commute with the projector to the flat bands), forbidding the representation $C_{2z}\overline{P} = \sigma_y \tau_x$. This is evidenced by the singular vales of $C_{2z}\overline{P}$ in Fig. \ref{fig6}(a) and Table \ref{tab3} (see also Refs. \cite{kang23,herzogarbeitman24}). Instead, we replace $C_{2z}\overline{P}$ by the unitary matrix $C_{2z}P = C_{2z}\overline{P}/\big(C_{2z} \overline{P} (C_{2z}\overline{P})^\dagger\big)^{1/2}$ and enforce Eq. (\ref{cherngauge}) for $C_{2z}P$. A similar procedure was applied when we assigned eigenvalues $\pm 1$ under $\sigma_z$ to the sublattice polarized states even though the eigenvalues of $\overline{\sigma}_z$ are different from $\pm 1$ (Fig. \ref{fig6}(a) and Table \ref{tab3}). Notice that because $C_{2z}P \neq C_{2z}\overline{P}$, $\{C_{2z}\overline{P}, \overline{\sigma}_z \}$ need not be equal to $0$ and $C_{2z}P$ might display components of the form $\tau_x$, $\tau_y$ and $\sigma_z\tau_y$ that cannot be removed with a gauge choice. Nevertheless, we find that these terms are negligible and we can satisfy Eq. (\ref{cherngauge}) up to numerical precision.

Two-fold rotations and time-reversal symmetries invert the momentum, $\kk \to -\kk$, and three-fold rotations rotate the momentum by $120^\circ$, $\kk \to C_{3z}(\kk)$, 
% and $C_{2x}$ reflects the $y$ component of the momentum, $\kk \to \text{diag}(1,-1)(\kk)$
up to translations by a dual lattice vector. We impose their action on the flavor degrees of freedom to be
\begin{align}
\begin{split}
    C_{2z} &= \begin{cases} \sigma_x \tau_x \quad &\kk = \Gamma_M \\
    -\sigma_x \tau_x \quad &\kk \neq \Gamma_M 
    \end{cases},  \\
     \mathcal{T} &= \begin{cases} \tau_x  \mathcal{K} \quad &\kk = \Gamma_M \\
    - \tau_x  \mathcal{K} \quad &\kk \neq \Gamma_M 
    \end{cases},  \\
    C_{3z} &= \begin{cases} \mathbb{1} \quad &\kk = \Gamma_M \\
    \text{exp}(-2\pi i/3 \ \sigma_z \tau_z) \quad &\kk \neq \Gamma_M 
    \end{cases}. 
    % \nonumber \\
    % C_{2x} &= \begin{cases} \sigma_x \quad &\kk = \Gamma_M \\
    % -\sigma_x &\kk \neq \Gamma_M 
    % \end{cases} \nonumber \\
\label{symreps}
\end{split}
\end{align}

The departure at $\Gamma_M$ from a 'uniform' representation can be traced back to the topological nature of the Chern bands. 
% In the language of the Topological Heavy Fermion model, the basis at $\Gamma_M$ is spanned by the $\Gamma_1 \oplus \Gamma_2$ topological electrons and away from $\Gamma_M$ by the local fermions, these orbitals having different symmetry properties. 
 
As is, the Chern gauge defined by Eqs. (\ref{cherngauge}),(\ref{symreps}) exhibits an ambiguity consisting of rotations of the states with the same Chern number, $\text{exp}(i\phi(\kk)\sigma_z\tau_z)$. We fix this phase by requiring that the Wannier functions obtained from the Chern bands are maximally localized \cite{xie2024chernbandsoptimallylocalized,gunawardana24}, following the algorithm of Ref. \cite{xie2024chernbandsoptimallylocalized}. By $C_{2z}\mathcal{T}$, the maximal localization procedure yields the same phase $\phi_K(\kk)$ for the $KA$ and $KB$ polarized bands, and $\phi_{K'}(\kk)$ for the $K'A$ and $K'B$ bands. Given that there exists some particle-hole asymmetry, i.e. $C_{2z}\overline{P} \neq C_{2z}P$, $\phi_K(\kk)$ need not be equal to $\phi_{K'}(\kk)$. In practice we find that they differ by at most $1^\circ$. We allow for this small error and take $\phi(\kk)$ to be the average. With this gauge transformation, the Bloch states are periodic and smooth except at the $\Gamma_M$ point, where the Berry connection acquires a vortex with a winding number equal to the Chern number \cite{xie2024chernbandsoptimallylocalized,gunawardana24}. 

Finally, we fix a global phase by imposing $C_{2y}\mathcal{T}= \mathcal{K}$ at $\Gamma_M$; our 'vortex Chern gauge' is then defined up to a global sign of all wave functions.

As a final comment, we have found that the phase $\phi(\kk)$ satisfies $\phi(\kk) = \phi(-\kk)$ exactly, and $\phi(\kk) \approx \phi(C_{3z}(\kk))$ only approximately (we attribute this to having taken the average of $\phi_K(\kk)$ and $\phi_{K'}(\kk)$ before). Consequently, the action of $C_{3z}$ is slightly modified to $C_{3z} = \text{exp}(-\lambda(\kk) \sigma_z \tau_z)$ with $\lambda(\Gamma_M) = 0$, $\lambda(K_M)=\lambda(K_M') = -2\pi/3$ and $\lambda(\kk) \approx -2\pi/3$ elsewhere, with deviations form $-2\pi/3$ always smaller than $3^\circ$. 
% and $\phi(\kk) \approx \phi(\text{diag}(1,-1)(\kk))$

\clearpage
\section{Symmetry of MATBG}\label{appd}

In the following we study the internal (flavor) symmetries of MATBG. For alternative discussions as well as additional details, see Refs. \cite{Bultinck20,bernevig21,song22}.

In the spirit of the continuum models, we can decompose a generic wave function into valley components,
\begin{align}
    \langle \boldsymbol{r} | \psi \rangle = \sum_{\tau} e^{i \tau \boldsymbol{K}_{l}\cdot \boldsymbol{r}} f^\psi_{\tau \sigma l}(\boldsymbol{r}).
    \label{decomp}
\end{align}
The phases $e^{\pm i\boldsymbol{K}_{l}\cdot \boldsymbol{r}}$ oscillate rapidly on the graphene lattice, whereas the envelope functions $f_{\tau \sigma l}(\boldsymbol{r})$ vary slowly on the graphene lattice and depend on the valleys $\tau = \pm$ and on the  sublattice $\sigma=A,B$ and layer $l= \text{top}\ (t),\text{bottom} \ (b)$ of the atom at $\boldsymbol{r}$. $(-)\boldsymbol{K}_l$ is the $K \ (K')$ point of layer $l$, see Fig. \ref{fig5}. We assume that $| \psi \rangle$ has a definite spin projection that is left implicit. This decomposition is only justified for states at low energies coming from the valleys of the parent graphene monolayers. In other words, the decomposition can be written only in effective models for MATBG where remote bands above a certain cutoff have been integrated out. In continuous models, the functions $f$ defined at the atomic positions are promoted to smooth functions in the continuum.

% Notice also that the phases $e^{\pm i\boldsymbol{K}_{l}\cdot \boldsymbol{r}}$ are not periodic on the moiré unit cell, $e^{i\boldsymbol{K_{t(b)}} \cdot \boldsymbol{L_1}}= e^{-(+)2\pi i/3}$ and $e^{i \boldsymbol{K_{t(b)}} \cdot \boldsymbol{L_2}} = e^{+(-)2\pi i/3}$ . Hence, in Bloch periodic functions $\langle \boldsymbol{r} + \boldsymbol{L}_{1,2} | \psi \rangle = e^{i \boldsymbol{k}\cdot \boldsymbol{L_{1,2}}}\langle \boldsymbol{r} | \psi \rangle $, the envelope functions must carry the Bloch phase plus an additional phase that cancels that of the oscillatory part.

\subsection*{Symmetry of ${\mathcal{H}}_{\text{int}}$}

Making use of Eq. (\ref{decomp}), consider the matrix element of $\mathcal{H}_\text{int}$ for generic states $\alpha ,\beta ,\gamma, \delta$ with valleys $\eta_\alpha, \eta_\beta, \eta_\gamma, \eta_\delta$,
\begin{align}
    {H}_{\text{int},\alpha \beta \gamma \delta} = & \sum_{\boldsymbol{r}\neq \boldsymbol{r'}} V(\boldsymbol{r}- \boldsymbol{r'})  e^{i (\tau_\delta -\tau_\alpha )\boldsymbol{K}_{l} \cdot \boldsymbol{r}} e^{i (\tau_\gamma -\tau_\beta )\boldsymbol{K}_{l'} \cdot \boldsymbol{r'}} f^\alpha_{\tau_\alpha \sigma l}(\boldsymbol{r})^* f^\delta_{\tau_\delta \sigma l}(\boldsymbol{r}) f^\beta_{\tau_\beta \sigma' l'}(\boldsymbol{r'})^* f^\gamma_{\tau_\gamma \sigma' l'}(\boldsymbol{r'}) \nonumber \\
    & + U \sum_{\boldsymbol{r}} e^{i (\tau_\delta + \tau_\gamma - \tau_\alpha - \tau_\beta)\boldsymbol{K}_{l} \cdot \boldsymbol{r}} f^\alpha_{\tau_\alpha \sigma l}(\boldsymbol{r})^* f^\delta_{\tau_\delta \sigma l}(\boldsymbol{r}) f^\beta_{\tau_\beta \sigma l}(\boldsymbol{r})^* f^\gamma_{\tau_\gamma \sigma l}(\boldsymbol{r}).
    \label{melem}
\end{align}
The nature of the Coulomb interaction forces the implicit spin components $s_\alpha = s_\delta$, $s_\beta = s_\gamma$ for both the long range and Hubbard terms, and $s_\alpha =  \Bar{s}_\beta $ for the Hubbard term. Notice also that we have omitted an essentially trivial piece coming from the $-\frac{1}{2}$ subtraction, see Eq. (\ref{hamiltoniansm}).

Clearly, the matrix element of Eq. (\ref{melem}) is the same as ${\mathcal{H}}_{\text{int},\alpha \beta \gamma \delta}$ if the initial and final states belong to the central bands. Nonetheless, ${\mathcal{H}}_{\text{int}}$ includes also a nontrivial local potential
\begin{align}
    \mathcal{H}_{\text{int}} \supset \sum_{\boldsymbol{r}s} \pi(\boldsymbol{r})\rho_{\boldsymbol{r}s}, \quad  \quad  \quad \quad \quad \pi(\boldsymbol{r}) = -2 \sum_{\boldsymbol{r'}\neq \boldsymbol{r}} V(\boldsymbol{r}- \boldsymbol{r'}) \rho_{0 \boldsymbol{r'}} - U \rho_{0 \boldsymbol{r}},    
\end{align}
see Eqs. (\ref{projhamiltoniansm}), (\ref{projhamiltoniansm2}), that we ought to consider separately.

\paragraph*{Valley-spin symmetries.} Looking at the first line of Eq. (\ref{melem}), only if $\tau_\alpha = \tau_\delta$ and $\tau_\beta = \tau_\gamma$ the matrix elements is non vanishing. Indeed, if $\tau_\delta = - \tau_\alpha$, we can take the $\boldsymbol{r}$, $\boldsymbol{r} + \boldsymbol{a}_1$ and $\boldsymbol{r} + \boldsymbol{a}_2$ and compute the phases $e^{2i\tau_\delta \boldsymbol{K}_{l} \cdot \boldsymbol{r}}$, $e^{2i\tau_\delta \boldsymbol{K}_{l} \cdot (\boldsymbol{r} + \boldsymbol{a}_1/2)} = e^{2i\tau_\delta \boldsymbol{K}_{l} \cdot \boldsymbol{r}} e^{\pm 2\pi i /3}$ and $e^{2i\tau_\delta \boldsymbol{K}_{l}  \cdot (\boldsymbol{r} + \boldsymbol{a}_2/2)} = e^{2i\tau_\delta \boldsymbol{K}_{l} \cdot \boldsymbol{r}} e^{\mp 2 \pi i/3}$. Given that the $f$ functions and the potential $V(\boldsymbol{r})$ vary slowly on atomic distances, $V(\boldsymbol{r} + \boldsymbol{a}_{1,2}) \approx V(\boldsymbol{r})$, $f(\boldsymbol{r}+ \boldsymbol{a}_{1,2} ) \approx f(\boldsymbol{r})$, the sum over the corresponding three atoms in ${\mathcal{H}}_{\text{int},\alpha \beta \gamma \delta}$ (modulo lattice relaxation and twist) cancels by $1 + e^{\pm 2\pi i /3} + e^{\mp 2\pi i /3} = 0$. Similarly occurs if $\tau_\gamma = - \tau_\beta$. 

The matrix element then has the structure $\mathcal{H}_{\text{int},\alpha \beta \gamma \delta} \sim \delta_{\tau_\alpha \tau_\delta} \delta_{\tau_\beta \tau_\gamma}\delta_{s_\alpha s_\delta} \delta_{s_\beta s_\gamma}$ with some coefficient that does not depend on the spin. Thus, the long range part of ${\mathcal{H}}_\text{int}$ enjoys a $SU(2)_K \times SU(2)_{K'} \times U(1)_\text{V}$ symmetry. The two $SU(2)$ correspond to independent spin rotations in each valley, and $U(1)_\text{V}$ to the valley charge conservation enforcing $\tau_\alpha + \tau_\beta = \tau_\gamma + \tau_\delta$. A heuristic explanation for the $SU(2)_K \times SU(2)_{K'}$ symmetry is that different valleys only interact via the total density in the Hartree channel, and because valley-dependent spin rotations do not change the density from each valley, these generate a symmetry.

On the other hand, the potential $\pi(\boldsymbol{r})$, being smooth on atomic distances, cannot produce high-momentum transfer scatterings between the two valleys, and it is independent of spin. Consequently, the symmetry is exactly preserved. 

Combined with the $U(1)_{\text{C}}$ of electric charge conservation we have a $U(2)_K \times U(2)_{K'}$ group with the 8 generators
\begin{align}
    \big[ s_i, {\mathcal{H}}_\text{int} \big] =  \big[ s_i \tau_z , {\mathcal{H}}_\text{int} \big] = 0, \quad \quad \quad \quad \quad (i=0,x,y,z).
    \label{u2xu2}
\end{align}
The operators $s_i$ are the identity ($i=0$) and Pauli ($i=x,y,z$) spin operators.

With respect to the second line of Eq. (\ref{melem}), by a similar argument we deduce that valley charge is conserved, $\tau_\alpha + \tau_\beta = \tau_\gamma + \tau_\delta$. However, only global spin rotations preserve the Hubbard Hamiltonian, breaking $U(2)_K \times U(2)_{K'}$ down to the physical symmetry of $SU(2)_\text{spin} \times U(1)_{\text{V}}$. Furthermore, the long range interaction also breaks $U(2)_K \times U(2)_{K'}$. Indeed, in the argument presented previously we assumed that the the potential $V(\boldsymbol{r})$ is slowly varying on atomic distances. This is not true near $\boldsymbol{r} = 0$ where there is a steep $1/|\boldsymbol{r}|$ dependence. In turn, there are symmetry breaking components coming from small distances. $U(2)_K \times U(2)_{K'}$-breaking interactions are called 'intervalley Hund's' couplings.

The electron-phonon interaction generates antiferromagnetic Hund's couplings (i.e. stabilizing opposite spins on opposite valleys) that can be decisive in selecting the ground state \cite{blason22,shi2024moireopticalphononsdancing,wang2024electronphononcouplingtopological,wang24_2,angeli19,kwan2023electronphononcouplingcompetingkekule}. On the other hand, the long range interaction is Hund-antiferromagentic \cite{sánchez2023correlatedinsulatorsmagicangle} and the on-site Hubbard interaction, ferromagnetic \cite{wang24_2,Chatterjee22}. These energy scales are much smaller than the chiral and flat-breaking terms  discussed below; we leave their study for a future publication.

\paragraph*{Chiral symmetry.} Now, consider the 'chiral' operator $\tilde{\sigma}_z$ in Eq. (\ref{vslop}). The exponential $\text{exp}(i\varphi \tilde{\sigma}_z)$ attaches the phase $e^{i\varphi}$ to $f_{\tau A l}$ and $e^{-i\varphi}$ to $f_{\tau B l}$. Clearly, this is an exact symmetry of ${{H}}_{\text{int}}$ because each phase encounters the complex conjugate in the matrix element, Eq. (\ref{melem}).

Nonetheless, the mode given by $\text{exp}(i\varphi \tilde{\sigma}_z)|\alpha\rangle$ generically does not belong to the subspace of central bands (in other words, it is not possible in general to construct a perfectly sublattice polarized basis in the effective theory), spoiling the chiral symmetry in $\mathcal{H}_{\text{int}}$.

To overcome this problem, we redefine the sublattice operator on the central bands as $\sigma_z = \sum_{\alpha\beta} [\sigma_z]_{\alpha \beta} c^\dagger_\alpha c_\beta$, with the matrix $\sigma_z = \overline{\sigma}_z / \big( \overline{\sigma}_z \overline{\sigma}_z^\dagger\big)^{1/2}$ and $[\overline{\sigma}_z]_{\alpha \beta} = \langle \alpha | \tilde{\sigma}_z | \beta \rangle$. This $\sigma_z$ here is the same as $\sigma_z$ acting on the Chern basis defined below Eq. (\ref{vslop}), having the property that it is the unitary matrix closest to $\overline{\sigma}_z$ (in the norm $||M|| = \text{tr}(M^\dagger M)$). We can assess the discrepancy between $\overline{\sigma}_z$ and $\sigma_z$ via the singular values or, equivalently up to a sign, the eigenvalues of $\overline{\sigma}_z$, see Table (\ref{tab3}). Singular values equal to $1$ imply that $\sigma_z = \overline{\sigma}_z$. 

With respect to the local potential, in the perfectly sublattice polarized limit the symmetry is exact owing to $\sigma_z \rho_{\boldsymbol{r}s} \sigma_z = \rho_{\boldsymbol{r}s}$. However, from the fact that $\overline{\sigma}_z \neq \sigma_z$ there appear intersublattice scatterings introducing some amount of symmetry breaking.

In turn, the exact chiral symmetry of ${H}_\text{int}$ turns into an approximate symmetry in the low-energy theory,
\begin{align}
    \big[\sigma_z \tau_z, {\mathcal{H}}_{\text{int}} \big] \approx 0,
\end{align}
the degree of non-symmetry indicated by the sublattice polarization of the central bands (we have attached the factor $\tau_z$ of the exact valley charge conservation, without need to change the argument).

\paragraph*{$C_{2z}P$ symmetry.} Let us turn to the particle-hole operator, $C_{2z}\tilde{P}$. This is a single-particle operator, local in real space and defined by its action on the $f$ functions in Eq. (\ref{decomp}),
\begin{align}
    C_{2z}\tilde{P} = \sigma_x \tau_y \mu_y,    
\end{align}
with $\mu_y$ the Pauli $y$ matrix in layer space, the top (bottom) layer having eigenvalue $1 \ (-1)$ under $\mu_z$. This operator is only properly defined on the continuous wave functions, but not in the atomistic theory. In practice, we obtain the smooth $f$ functions by interpolation in a valley polarized basis, and later they are sampled in the atomic positions of the opposite sublattice and layer. The inner products of the transformed wave functions with the basis wave functions define the matrix elements of $C_{2z}\overline{P}$.

We will argue now that the matrix element ${{H}}_{\text{int},\alpha' \beta' \gamma' \delta'}$ between modes $\eta' = C_{2z}\tilde{P}|\eta\rangle$ ($\eta=\alpha,\beta,\gamma,\delta$) is approximately equal to ${{H}}_{\text{int}, \alpha \beta \gamma \delta}$. For that, we may establish a one-to-one correspondence between atoms $\boldsymbol{r}$ and $\boldsymbol{s}$, with $\boldsymbol{s}$ belonging to the opposite sublattice and layer to $\boldsymbol{r}$ and with approximately the same in-plane coordinates. Then, we can replace the sums over $\boldsymbol{r},\boldsymbol{r'}$ in Eq. (\ref{melem}) by sums over $\boldsymbol{s},\boldsymbol{s'}$ and conclude that, in the $U(2)_{K} \times U(2)_{K'}$ limit where $\tau_\alpha=\tau_\delta$, $\tau_\beta=\tau_\gamma$ and the valley phases cancel, ${{H}}_{\text{int}, \alpha \beta \gamma \delta} = {{H}}_{\text{int}, \alpha' \beta' \gamma' \delta'}$ to a very good approximation.
This property hints to the existence of a continuous symmetry generated by $C_{2z}\tilde{P}$. Indeed, in Appendix D of Ref. \cite{sanchez24} we showed that $[{H}_\text{int}, C_{2z}\tilde{P}] = 0$ (assuming $U(2)_{K} \times U(2)_{K'}$ and up to effects from the discrete lattice), also we provide more details on the implementation of $C_{2z}\tilde{P}$ in the atomistic model in Appendix C of Ref. \cite{sanchez24}. 

Again, the symmetry is not well defined in the effective theory hence, analogously to $\sigma_z$, we define $C_{2z}P = \sum_{\alpha\beta} [C_{2z}P]_{\alpha \beta} c^\dagger_\alpha c_\beta$, with the matrix $C_{2z}P = C_{2z}\overline{P} /\big(C_{2z} \overline{P} (C_{2z}\overline{P})^\dagger\big)^{1/2}$ and $[C_{2z}\overline{P}]_{\alpha \beta} = \langle \alpha |C_{2z}\tilde{P} | \beta \rangle$.  

\begin{figure}[h!]
    \centering
    \includegraphics[width=.9\linewidth]{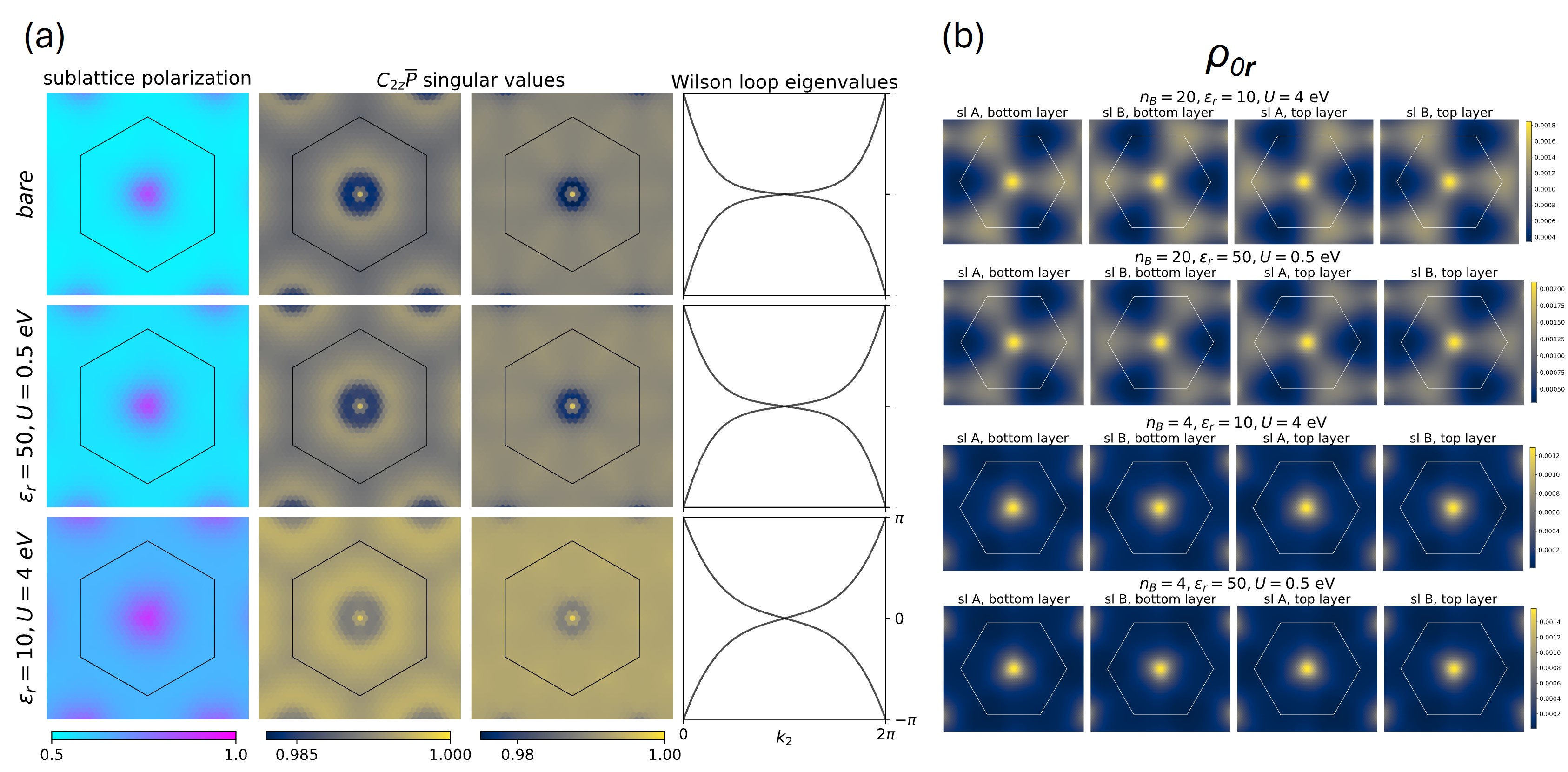}
    \caption{(a) Hilbert space of the bare and renormalized flat manifold. In the first column we show the positive eigenvalue of the sublattice operator, $\overline{\sigma}_z$, (sublattice polarization) for the valley $K$ sector on the Brillouin zone. The relations $\{C_{2z}\mathcal{T}, \overline{\sigma}_z \} = 0$ and $[\mathcal{T}, \overline{\sigma}_z ] = 0$ constrain the negative eigenvalue and the eigenvalues for valley $K'$, respectively. In the second and third columns we plot the two distinct singular values of the particle-hole operator $C_{2z}\overline{P}$, which are degenerate in pairs by $C_{2z}\overline{P} = (C_{2z}\overline{P})^\dagger$ and $\{C_{2z}\overline{P},\tau_z\}=0$. In the fourth column we plot the phase of the eigenvalues of the Wilson loop operator \cite{song21,alexandrinata14} for the valley $K$ sector. (b) Density of the renormalized symmetric state, $\rho_{0\boldsymbol{r}}$, for $n_B=4,20$ and $(\epsilon_r, U) = (10, 4 \text{ eV})$, $(50, 0.5 \text{ eV})$; 'sl' is short for sublattice. The densities on opposite sublattice and layers are approximately equal.}
    \label{fig6}
\end{figure}

With regards to the local potential, in the case of perfect particle-hole symmetry, and neglecting intervalley scatterings, we have $C_{2z}{P} \rho_{\boldsymbol{r}s} C_{2z}{P} = \rho_{\boldsymbol{s}s}$. Then, provided $\rho_{0\boldsymbol{r}} = \rho_{0\boldsymbol{s}}$, the potential respects the $C_{2z}P$ symmetry. We find that $\rho_{0 \boldsymbol{r}} \approx \rho_{0\boldsymbol{s}}$ to a good approximation, as shown in Fig. \ref{fig6}(b); moreover we also find that $\rho_{0\boldsymbol{r}} \approx \frac{1}{2}\sum_{\boldsymbol{k}n}' |\langle \boldsymbol{r} | \boldsymbol{k} n \rangle |^2$. These properties of $\rho_{0\boldsymbol{r}}$ can be partly understood from the constrains imposed by the crystallographic and valley symmetries and the properties of the normal state under $C_{2z}P$ and $\sigma_z$, see Eq. (\ref{h0sym}) (where we can replace $\mathcal{H}_0$ by the normal state, $H_{\text{MF}}$). Notice also that in $\pi(\boldsymbol{r})$ the density is convolved with the Coulomb potential, which tends to smoothen the inhomogeneities making the relation $\pi(\boldsymbol{r}) \approx \pi(\boldsymbol{s})$ tighter. Particle-hole asymmetry, $C_{2z}\overline{P} \neq C_{2z}P$, translates to some amount of symmetry breaking.

We conclude that $C_{2z}P$ generates an approximate symmetry in the low-energy theory,
\begin{align}
    \big[  C_{2z}P, {\mathcal{H}}_\text{int}\big] \approx 0,  
\end{align}
and the strength of symmetry breaking can be assessed by the singular values of $C_{2z}\overline{P}$, displayed in Fig. \ref{fig6}(a) and Table (\ref{tab3}). 

\paragraph*{$R$ symmetry.} Consider as well the product of operators
\begin{align}
    \tilde{R} = C_{2z}\tilde{P}\tilde{\sigma}_z\tau_z = \sigma_y \tau_x \mu_y,
\end{align}
which is clearly conserved by the interaction, $[{H}_\text{int}, \tilde{R}] = 0$, being the product of exact symmetries. Again, the symmetry becomes approximate in the effective theory with generator $R$, 
\begin{align}
    \big[  R, {\mathcal{H}}_\text{int} \big] \approx 0,
\end{align}
and the strength of the symmetry breaking can be deduced from the singular values of $\overline{R}$. Here we have defined $R$ and $\overline{R}$ analogously to $\overline{\sigma}_z$ and $C_{2z}\overline{P}$, and $\sigma_z$ and $C_{2z}P$, respectively.

It is important to note that $\overline{R}$ need not be equal to $C_{2z}\overline{P}\overline{\sigma}_z\tau_z$; the projection of $\tilde{R} = C_{2z}\tilde{P}\tilde{\sigma}_z\tau_z$ is not the same as the product of the projections of $ C_{2z}\tilde{P}$ and $\tilde{\sigma}_z\tau_z$. Nonetheless, the low-energy Hilbert space is very particle-hole symmetric (the singular values of $C_{2z}\overline{P}$ in the flat bands are of the order of $0.99$), thus we conclude that in fact $\overline{R} \approx C_{2z}\overline{P}\overline{\sigma}_z\tau_z$. Consequently, the singular values of $\overline{R}$ will be  approximately equal to the singular values of $\overline{\sigma}_z$, of the order of $0.5-0.9$ in the flat bands.

By the same reason, the unitary matrix $R$ is not necessarily equal to $C_{2z}{P}  \sigma_z \tau_z$. For instance, in the flat-band theory with the Chern gauge, the relations $\{R,\tau_z\} = \{R, C_{2z}\mathcal{T}\} = \{R, \tilde{\sigma}_z \} = 0$ impose the form $R = \sin(\nu(\kk)) \sigma_y\tau_y + \cos(\nu(\kk)) \sigma_x\tau_y$ for some momentum-dependent angle $\nu(\kk)$. Observe that the property $[C_{2z}\tilde{P},\tilde{R}] = 0$ no longer holds for $[C_{2z}P,R]$. Again, almost particle-hole symmetry obliges $R \approx$ $C_{2z}{P}  \sigma_z \tau_z = \sigma_x\tau_y$, i.e. $\nu(\kk) \approx 0$. Indeed, $\nu(\kk)$ is at most $\pm 1.2^\circ$. We ignore such small deviations from $\nu(\kk) = 0$ and consider the symmetry as generated by $R = C_{2z}{P}  \sigma_z \tau_z=\sigma_x\tau_y$.

% \begin{figure}
%     \centering
%     \includegraphics[width=0.9\linewidth]{n0r.jpg}
%     \caption{Density of the renormalized symmetric state, $\overline{n}_{0\boldsymbol{r}}$, for $n_B=4,20$ and $(\epsilon_r, U) = (10, 4 \text{eV}), (50, 0.5 \text{eV})$. Opposite sublattices and layers display approximately the same profile.}
%     \label{fig6}
% \end{figure}

\subsection*{Full symmetry group: chiral and flat limits}

Once identified the 'hidden' symmetries, we can obtain the full symmetry group by commuting the $U(2)_K \times U(2)_{K'}$ generators of Eq. (\ref{u2xu2}) with $C_{2z}P$, $R$ and/or $\sigma_z \tau_z$. For simplicity, let us consider the spinless system, in which $U(2)_K \times U(2)_{K'}$ turns to $U(1)_\text{C} \times U(1)_\text{V}$, generated by the identity, $\mathbb{1}$, and $\tau_z$. Let us also work in the Chern basis of the $n_B=4$ model for concreteness. The symmetry generators are written as $\mathcal{G} = \sum_{\kk} c^\dagger_{\kk} G c_{\kk}$ with $c^\dagger_{\kk} = \big(c^\dagger_{\kk \alpha}\big)$ a vector and $G=[G]_{\alpha \beta}$ a matrix in the Chern basis; let us work with just the $G$ matrices for neatness. 

With $C_{2z}P$, we can obtain $S_{x} = \frac{-i}{2}[\tau_z,C_{2z}P] = \sigma_x\tau_x$. Denoting $S_0 = \mathbb{1}$, $S_y = C_{2z}P$ and  $S_z = \tau_z$ we see that $\{S_0,S_x,S_y,S_z\}$ generates a $U(2)$ group. On the other hand, with the $R$ symmetry we get a different $U(2)$ group with $S'_0 = \mathbb{1}$, $S'_x = R$, $S'_{y} = \frac{i}{2}[\tau_z,R] = \sigma_y \tau_y$ and $S'_z = \tau_z$ (the $\sigma_z\tau_z$ symmetry results unphysical, spoiled by both $\mathcal{H}_{\text{int}}$ and $\mathcal{H}_0$, and we avoid discussing it). In the limit with both $C_{2z}P$ and $R$ (hence also $\sigma_z\tau_z$), there is a total of $8$ generators, $\{\mathbb{1},\sigma_x\tau_x,\sigma_x\tau_y,\sigma_y\tau_x\sigma_y\tau_y,\tau_z,\sigma_z\tau_z \}$. One can check that an alternative basis of the Lie algebra is $P_{\pm} \{\mathbb{1}, \sigma_x\tau_x, \sigma_x\tau_y, \tau_z \} P_{\pm}$, with $P_{\pm} = 1/2(\mathbb{1}\pm \sigma_z \tau_z)$ the projector to the Chern $\pm$ sector. It becomes clear that this is the Lie algebra of a $U(2) \times U(2)$ group of independent $U(2)$ transformations in each Chern sector. 

The same procedure can be followed straightforwardly including the spin; the $U(2)$ groups are enlarged to $U(4)$ and the $U(2)\times U(2)$ to $U(4)\times U(4)$. The different limits and their symmetry groups are listed in Table (\ref{tab2}). The case with $U(2)_K \times U(2)_{K'}$ and $C_{2z}P$ is dubbed 'flat' limit and the symmetry $U(4)_\text{flat}$ (the dispersion $\mathcal{H}_0$ is prone to breaking this symmetry, as we discuss below). On the other hand, with $U(2)_K \times U(2)_{K'}$ and $R$ we are in the 'chiral-nonflat' limit and the symmetry is denoted $U(4)_\text{chiral}$. Note that in the main text we use the notations $C_{2z}P = \mathcal{G}_{\text{flat}}$ and $R= \mathcal{G}_{\text{chiral}}$.

% \begin{align}
%     U(4) \times U(4):& \quad \quad G_n = \{t_a,t_a\sigma_x\tau_y, t_a\sigma_y\tau_x, t_a\sigma_z\tau_z\} \nonumber \\
%     U(4)_{\text{chiral-nonflat}}:& \quad \quad G_n = \{t_a,t_a\sigma_y\tau_x \} \nonumber \\
%     U(4)_{\text{flat}}:& \quad \quad G_n = \{t_a,t_a\sigma_x\tau_y\} \nonumber \\
%     U(2)_K \times U(2)_{K'}: & \quad \quad G_n = \{t_a \} = \{ s_i, s_i\tau_z\} \quad (i=0,x,y,z) \nonumber \\
%     \text{symmetry generators}& \quad \mathcal{G}_n = \sum_{\kk} c^\dagger_{\kk} G_n c_{\kk} \nonumber \\
% \end{align}

\begin{table}[h!]
    \centering
    \renewcommand{\arraystretch}{1.25} % Default value: 1
    \begin{tabular}{ccc}
        \hline \hline
        symmetry & group & generators $\mathcal{G} = \sum_{\kk}
        c^\dagger_{\kk} G c_{\kk}$ \\ 
        \hline  
        chiral-flat & $U(4) \times U(4)$ & $G = \{t_a,t_a\sigma_x\tau_y, t_a\sigma_y\tau_x, t_a\sigma_z\}$ \\ 
        % \hline
        chiral-nonflat & $U(4)_{\text{chiral}}$ & $G = \{t_a,t_a\sigma_y\tau_x \}$ \\ 
        % \hline
        nonchiral-flat & $U(4)_{\text{flat}}$ &  $G = \{t_a,t_a\sigma_x\tau_y\}$ \\ 
        % \hline
        nonchiral-nonflat & $U(2)_K \times U(2)_{K'}$ & $G = \{t_a \} = \{ s_i, s_i\tau_z\} \quad (i=0,x,y,z)$ \\ 
        physical & $SU(2)_{\text{spin}} \times U(1)_{\text{V}} \times U(1)_\text{C}$ & $G = \{ s_i,\tau_z, \mathbb{1} \} \quad (i=x,y,z)$ \\ 
        \hline \hline
    \end{tabular}
    \caption{Different limits and their symmetry groups in the Chern basis of $n_B=4$ MATBG. The explicit symmetry breaking hierarchy is $U(4) \times U(4) \to U(4)_{\text{chiral}} \to U(2)_K \times U(2)_{K'} \to SU(2)_{\text{spin}} \times U(1)_{\text{V}} \times U(1)_\text{C}$ as we have discussed in this work.}
    \label{tab2}
\end{table}

\subsection*{Symmetry of $\mathcal{H}_0$}

% Let us investigate the symmetry of the dispersion, $\mathcal{H}_0$. 
First of all, let us briefly review the BM and THF models. In these models we have $\{H_0,C_{2z}\tilde{P}\} = 0$, and in the chiral limit, $\{H_0,\tilde{\sigma}_z \tau_z\} = [H_0,\tilde{R}] = 0$ ($H_0$ is the bare dispersion of the BM or THF model; in the THF model, a modified sublattice operator similar to our $\sigma_z$ is used).

We see that only in the chiral limit the $\tilde{R}$ symmetry remains. As a matter of fact, the relations $\{H_0,C_{2z}\tilde{P}\} = 0$ and (approximately) $\{H_0,\tilde{\sigma}_z \tau_z\} = 0$ imply that the symmetry is broken by energy scales as large as the full bandwidth of the theory. Projecting to the flat bands effectively reduces the bandwidth and in strong coupling, $H_\text{int} \gg H_0 \approx 0$, we have $[C_{2z}\tilde{P},H_0] = [C_{2z}\tilde{P},H_\text{int}] = 0$ and additionally $[\tilde{\sigma}_z\tau_z,H_0] = [\tilde{\sigma}_z\tau_z,H_\text{int}] = 0$ in the chiral limit. Nonetheless, a proper many-body projection to the flat bands will presumably induce new components that widen the bands of $H_0$ at the bare magic angle as we have shown in this work, disallowing the approximation $H_0 \approx 0$ (at the renormalized magic angle the induce terms will tend to cancel the bare $H_0$ thus recovering the strong coupling limit).

Back to our system, in the normal state intervalley scattering is negligible and the spins are identical, so the $U(2)_K \times U(2)_{K'}$ group is exactly preserved by the dispersion. For the $C_{2z}P$, $R$ and $\sigma_z\tau_z$ symmetries, let us focus on the $n_B=4$ theory. As stated in the main text, we can write ${h}(\kk)$ in the vortex Chern basis and decompose it into real components,
% \begin{gather}
\begin{align}
    {h}(\kk) &=  {h}_{00}(\kk) + {h}_{z0}(\kk)\tau_z + {h}_{x0}(\kk)\sigma_x + {h}_{yz}(\kk)\sigma_y\tau_z + {h}_{y0}(\kk)\sigma_y + {h}_{xz}(\kk)\sigma_x\tau_z, 
    \label{hk}
% \end{gather}
\end{align}

The strength of the different components is summarized in Table (\ref{tab3}). The dominant term is $h_{x0}(\kk) \sigma_x + h_{yz}(\kk) \sigma_y \tau_z$, with ${h}_{00}(\kk)$ also sizable at the $\Gamma_M$ point. In fact, $h_{x0}(\kk)$ and $h_{yz}(\kk)$ are the main sources of symmetry breaking as we discuss in the main text. We have then the relations (neglecting $h_{00}$)
\label{h0sym}
\begin{align}
\big\{ C_{2z}P, \mathcal{H}_0 \big\} \approx 0, \quad \quad \quad
 \big\{\sigma_z\tau_z, \mathcal{H}_0 \big\} \approx 0, \quad \quad \quad
  \big[R, \mathcal{H}_0  \big] \approx 0.
  \label{hcomms}
\end{align}
Notice that the property of the chiral limit, $\{\sigma_z\tau_z, \mathcal{H}_0 \} \approx 0$, is realized to a good approximation in $\mathcal{H}_0$. The breaking of the chiral symmetry in our model occurs in ${\mathcal{H}}_{\text{int}}$ owing to $\overline{\sigma}_z \neq \sigma_z$ as we discussed above.

Then, $\mathcal{H}_0$ breaks the $C_{2z}P$ and $\sigma_z\tau_z$ symmetry. All the more, in theories with larger bandwidth, say $n_B=20$, and assuming that the relations of Eq. (\ref{hcomms}) are approximately satisfied, the $C_{2z}P$ and $\sigma_z\tau_z$ symmetries are more seriously broken. In this respect, the authors of Ref. \cite{song22} discovered a chiral operator having a simple form in the THF basis, denoted $\tilde{S}$, with $[ \tilde{S}, H_{\text{int}} ] \approx 0$ and $\{\tilde{S}, H_0 \} \approx 0$; the non-anticommuting terms being of the order of the flat bandwidth. This 'third chiral' operator can be combined with the particle-hole operator resulting in $[C_{2z}\tilde{P}\tilde{S},H_0] \approx 0$ and $[C_{2z}\tilde{P}\tilde{S},H_\text{int}] \approx 0$. It was shown that $\tilde{S}$ is consistent with the identity in (most of) the flat bands, resulting in that $C_{2z}\tilde{P}\tilde{S}$ approximately anticommutes with the flat bands and approximately commutes with the higher bands of $H_0$. This is in contrast to $C_{2z}\tilde{P}$ that anticommutes with all the bands. Similarly, the operator $\tilde{\sigma}_z\tau_z \tilde{S}$ approximately commutes with $H_\text{int}$ and with the high-energy bands of $H_0$ (even though the relation $\{\tilde{\sigma}_z,H_0 \} \approx 0$ is expected to become worse on the higher bands), and approximately anticommutes with the flat bands.

We anticipate that the third chiral operator $\tilde{S}$ of Ref. \cite{song22} will lead to analogous relations for the operators $C_{2z}PS$, $\sigma_z\tau_z S$, ${\mathcal{H}}_\text{int}$, $\mathcal{H}_0$ of the projected theory, effectively replacing $\sigma_{z}\tau_z$ and $C_{2z}P$ by $C_{2z}PS$ and $\sigma_{z}\tau_z S$ as the generators of the approximate symmetries.

\begin{table}[h!]
    \centering
    \renewcommand{\arraystretch}{1.25} % Default value: 1
    \begin{tabular}{cccc}
        \hline \hline
         & bare ($\epsilon_r^{-1}=0$, $U=0$ eV) & $\epsilon_r =50$, $U=0.5$ eV & $\epsilon_r =10$, $U=4$ eV  \\ \hline
        minimal  $\overline{\sigma}_z$ eigenvalue  & $0.524$ & $0.548$ & $0.638$ \\  
        % \hline
        maximal  $\overline{\sigma}_z$ eigenvalue  & $0.849$ & $0.858$ & $0.890$\\  \
        % \hline
        minimal  $C_{2z}\overline{P}$ singular value  & $0.975$ & $0.978$ & $0.988$\\ 
        % \hline 
        maximal  $C_{2z}\overline{P}$ singular value  & $0.996$ & $0.997$ & $0.998$ \\  
        % \hline 
        av$_{BZ}$  $|{h}_{0z}|$ (meV) & $0.16$ & $0.27$ & $0.76$\\  
        % \hline 
        av$_{BZ}$  $|{h}_{y0} + i{h}_{xz}|$ (meV) & $0.35$ & $0.39$  & $0.43$\\  
        % \hline 
        av$_{BZ}$  $| {h}_{x0} + i{h}_{yz}|$ (meV) & $0.64$  & $1.36$ & $5.85$\\  
        % \hline 
        $| {h }_{x0}(\Gamma_M) + i{h}_{yz}(\Gamma_M)|$ (meV) & $4.34$ & $7.04$ & $20.99$ \\
        % \hline 
        av$_{BZ}$ ${h}_{00}$ (meV) & $0.54$  & $0.69$ & $1.23$ \\ 
        % \hline 
        ${h}_{00}(\Gamma_M)$ (meV) & $5.79$ & $6.45$ & $8.21$ \\ 
        \hline \hline
        \end{tabular}
    \caption{Relevant figures for the breaking of the $U(4) \times U(4)$ symmetry in the $n_B=4$ system. Singular values of $C_{2z}\overline{P}$ different from $1$ break the flat symmetry of $\mathcal{H}_{\text{int}}$, and eigenvalues of $\overline{\sigma}_z$ different from $1$ break the chiral symmetry of $\mathcal{H}_{\text{int}}$. The components $h_{0z}(\kk)$, $h_{x0}(\kk)$ and $h_{yz}(\kk)$ break the flat symmetry of $\mathcal{H}_0$, and the components $h_{0z}(\kk)$, $h_{y0}(\kk)$ and $h_{xz}(\kk)$ break the chiral symmetry of $\mathcal{H}_0$. av$_{BZ}$ denotes the average over the Brillouin zone.}
    \label{tab3}
\end{table}

% \clearpage
\section{Topological phase transition in the graphene subtraction scheme}\label{appe}

It has been argued that the tight-binding function $t(\boldsymbol{r})$ is derived in such a way that some long range exchange is already taken into account, so one must subtract a counter-term to the interaction in order to avoid double counting. In the graphene subtraction scheme, the Hartree-Fock potential of two decoupled graphene layers at neutrality is chosen for subtraction. In other words, $H_\text{int}$ is normal-ordered \cite{Giuliani_Vignale_2005} with respect to the state of two decoupled graphene layers at the neutrality point. The Hamiltonian then reads
\begin{alignat}{2}
    &{H} =  H_{\text{TB}} + {H}_\text{int}  \\ 
    &H_{\text{TB}} = \sum_{\boldsymbol{r}\boldsymbol{r'}s} t(\boldsymbol{r}-\boldsymbol{r'}) c^\dagger_{\boldsymbol{r}s}c_{\boldsymbol{r'}s}   \\  
    &{H}_\text{int} = \frac{1}{2}\sum_{\boldsymbol{r}\neq \boldsymbol{r'}ss'} V(\boldsymbol{r}- \boldsymbol{r'}) :c^\dagger_{\boldsymbol{r}s}  c_{\boldsymbol{r}s}c^\dagger_{\boldsymbol{r'}s'}  c_{\boldsymbol{r'}s'}: + \sum_{\boldsymbol{r}} U :c^\dagger_{\boldsymbol{r}\uparrow}  c_{\boldsymbol{r}\uparrow}c^\dagger_{\boldsymbol{r}\downarrow}  c_{\boldsymbol{r}\downarrow}: \nonumber \\
    &\hspace{.67cm} = \frac{1}{2}\sum_{\boldsymbol{r}\neq \boldsymbol{r'}ss'} V(\boldsymbol{r}- \boldsymbol{r'}) c^\dagger_{\boldsymbol{r}s}  c_{\boldsymbol{r}s}c^\dagger_{\boldsymbol{r'}s'}  c_{\boldsymbol{r'}s'} + \sum_{\boldsymbol{r}} U c^\dagger_{\boldsymbol{r}\uparrow}  c_{\boldsymbol{r}\uparrow}c^\dagger_{\boldsymbol{r}\downarrow}  c_{\boldsymbol{r}\downarrow} \nonumber \\
    &\hspace{1.05cm} -\sum_{\boldsymbol{r}\neq \boldsymbol{r'},ss'} V(\boldsymbol{r}- \boldsymbol{r'}) \langle c^\dagger_{\boldsymbol{r}s'} c_{\boldsymbol{r}s'} \rangle_{\text{gr}} \ c^\dagger_{\boldsymbol{r}s} c_{\boldsymbol{r}s}
    + \sum_{\boldsymbol{r}\neq \boldsymbol{r'},s} V(\boldsymbol{r}- \boldsymbol{r}) \langle c^\dagger_{\boldsymbol{r'}s} c_{\boldsymbol{r}s} \rangle_{\text{gr}} \ c^\dagger_{\boldsymbol{r}s} c_{\boldsymbol{r'}s}  \nonumber \\ 
    &\hspace{1.05cm} - \sum_{\boldsymbol{r} s} U  \langle c^\dagger_{\boldsymbol{r}\Bar{s}} c_{\boldsymbol{r}\Bar{s}} \rangle_{\text{gr}} \ c^\dagger_{\boldsymbol{r}s} c_{\boldsymbol{r'}s} + \text{constant},
\end{alignat}
with $\langle ... \rangle_{\text{gr}}$ the expectation value in the state of the decoupled layers of graphene. The mean-field Hamiltonian reads
\begin{align}
    {H}_{\text{MF}} =& \sum_{\boldsymbol{r}\boldsymbol{r'}s} t(\boldsymbol{r} - \boldsymbol{r'}) c^\dagger_{\boldsymbol{r}s}c_{\boldsymbol{r'}s} + \sum_{\boldsymbol{r}\neq \boldsymbol{r'},ss'} V(\boldsymbol{r}- \boldsymbol{r'}) \bigg(\langle c^\dagger_{\boldsymbol{r}s'} c_{\boldsymbol{r}s'} \rangle_0  - \langle c^\dagger_{\boldsymbol{r}s'} c_{\boldsymbol{r}s'} \rangle_{\text{gr}} \bigg) c^\dagger_{\boldsymbol{r}s} c_{\boldsymbol{r}s} \nonumber \\ 
    &-\sum_{\boldsymbol{r}\neq \boldsymbol{r'},s} V(\boldsymbol{r}- \boldsymbol{r})  \bigg( \langle c^\dagger_{\boldsymbol{r'}s} c_{\boldsymbol{r}s} \rangle_0 - \langle c^\dagger_{\boldsymbol{r'}s} c_{\boldsymbol{r}s} \rangle_{\text{gr}} \bigg) c^\dagger_{\boldsymbol{r}s} c_{\boldsymbol{r'}s} + \sum_{\boldsymbol{r} s} U \bigg( \langle c^\dagger_{\boldsymbol{r}\Bar{s}} c_{\boldsymbol{r}\Bar{s}} \rangle_0  - \langle c^\dagger_{\boldsymbol{r}\Bar{s}} c_{\boldsymbol{r}\Bar{s}} \rangle_{\text{gr}}  \bigg) c^\dagger_{\boldsymbol{r}s} c_{\boldsymbol{r'}s} + \text{constant}.
\end{align}
It was assumed that the contributions to the Fock matrix $ \langle c^\dagger_{\boldsymbol{r'}s} c_{\boldsymbol{r}s} \rangle_0$ coming from high-energy states are essentially identical to those of the decoupled layers, contributing to $ \langle c^\dagger_{\boldsymbol{r'}s} c_{\boldsymbol{r}s} \rangle_{\text{gr}}$. Hence, the virtual processes with remote bands would be suppressed and the normal state will be close to the ground state of the tight-binding Hamiltonian.

\begin{figure*}[h!]
    \includegraphics[width=.22\linewidth]{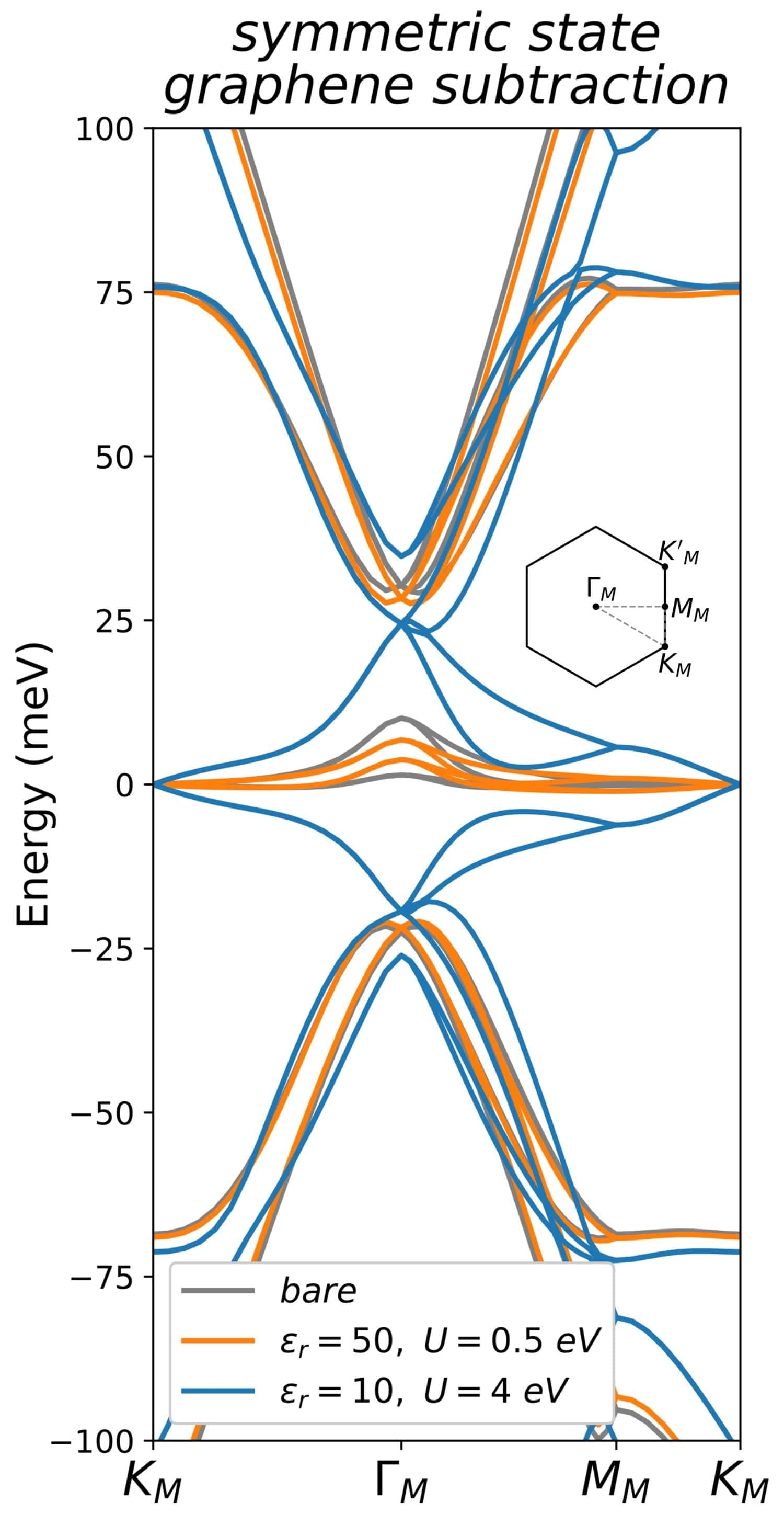}
    \caption{Band structures of the symmetric state in the graphene subtraction scheme for $\epsilon_r=10$, $U=4$ eV and $\epsilon_r=50$, $U=0.5$ eV, compared to the non-interacting (bare) bands. The system underwent a topological phase transition between $\epsilon_r=50$ and $10$ and the flat bands become doubly (per valley) degenerate at $\Gamma_M$. The inset shows the Brillouin zone and the momentum lines plotted.}
    \label{fig7}
\end{figure*}

Contrary to this expectation, our calculations with the graphene subtraction also display significant band widening, as depicted in Fig. \ref{fig7}. All the more, for some value of $\epsilon_r$ between $50$ and $10$ we find a topological phase transition moving the one-dimensional (two-dimensional) representations (per valley) at $\Gamma_M$ from the flat (remote) bands to the remote (flat) bands. Notice that the flat bands become entangled with the higher bands making the $n_B=4$ projector ill-defined. 

\clearpage
\section{Additional information: many-body projection}\label{appf}

The many-body projection introduced in Appendix \ref{appb} relies on the band projector onto excitations of the central $n_B$ bands, Eq. \ref{projwf}. This projector is well defined  if the set of central bands is gapped from the remaining bands, and it can be approximated numerically via a discretization of the Brillouin zone. In MATBG we can consider a weaker condition, which is that the central $n_B/2$ bands are gapped from the remaining bands for each valley, because the two valleys are completely decoupled
% , hence band crossings between different valleys do not carry a singular behavior of the wave functions.

\begin{figure*}[h!]
    \includegraphics[width=.85\linewidth]{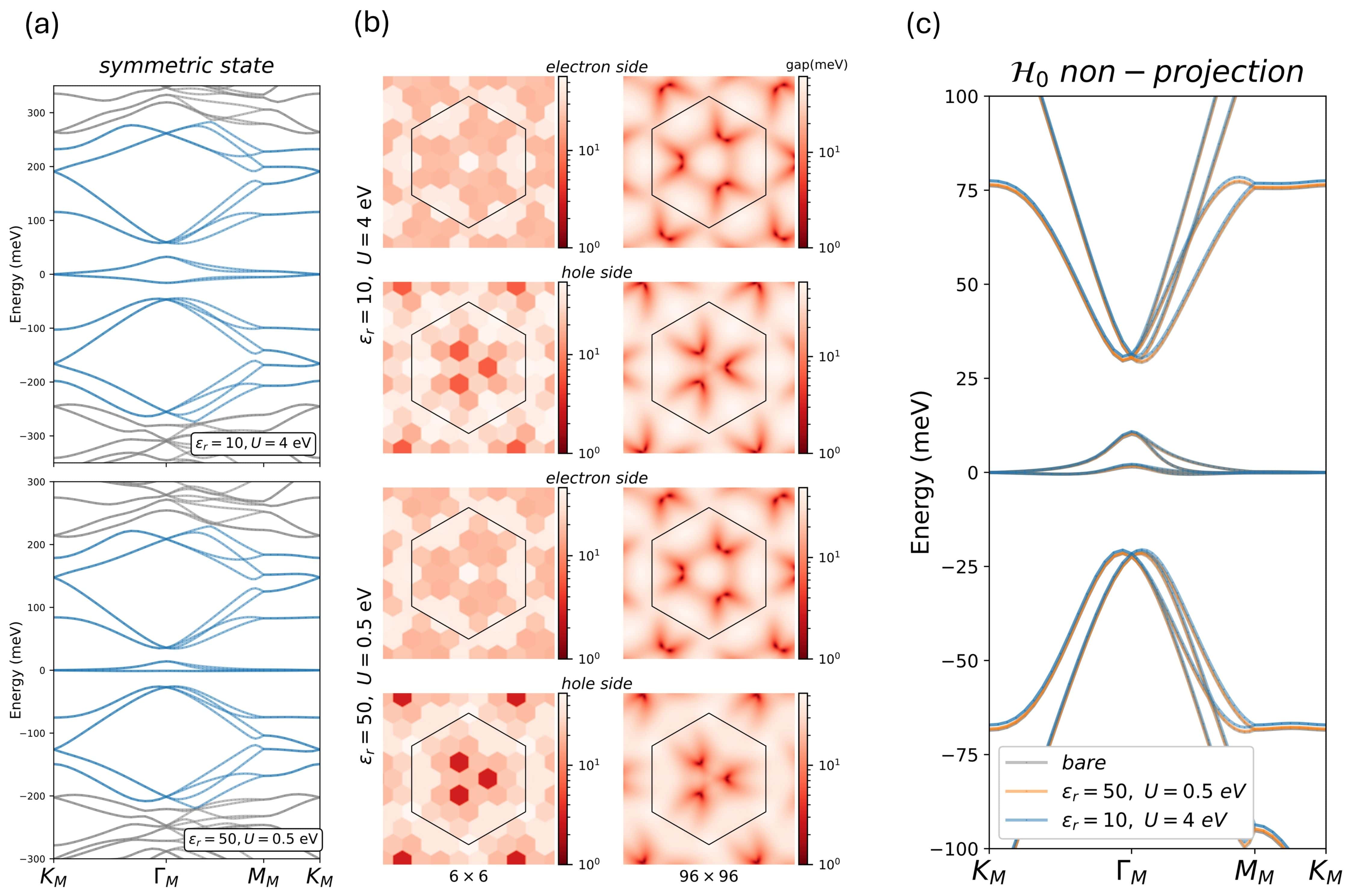}
    \caption{(a) Band structures of the symmetric state for $\epsilon_r=10$, $U=4$ eV and $\epsilon_r=50$, $U=0.5$ eV, showing the $20$ central bands in blue. (b) Gaps between the central and remote bands on the electron and hole side, for the states of valley $K$. On the left column there are the $6\times 6$ discretized Brillouin zone and on the right, the $96 \times 96$ zone showing the band touchings near the $\Gamma_M M_M$ or symmetry-related lines. (c) Band structures of $\mathcal{H}_0$ in the non-projected ($n_B=11908$) model, compared to the non-interacting (bare) bands. They are almost identical.}
    \label{fig8}
\end{figure*}

For the $n_B=4$ theory, the flat bands are clearly isolated from the remote bands, with gaps of some tens of meV. Let us focus then on $n_B=20$. In Fig. \ref{fig8}(a) we plot the band structures of the symmetric state along the high-symmetry $K_M \Gamma_M M_M K_M$ line; the lowest $20$ bands are shown in blue. Clearly, the set of the first $20$ bands is gapped from the remaining bands at the high symmetry points $\Gamma_M$, $K_M$ and $M_M$. Moreover, the symmetry representations are consistent with those of a topologically trivial system with ten orbitals per valley \cite{carr19,po19}.

In Fig. \ref{fig8}(b) we show the gaps on the electron and hole sides on the Brillouin zone. The values correspond to the valley $K$ bands, i.e. bands with positive $\tilde{\tau}_z$ expectation value (Eq. \ref{vslop}); valley $K'$ is obtained by a $180^\circ$ rotation. On the left column we consider the $6 \times 6$ discretization used in our self-consistent algorithm. As anticipated by the analysis of the high-symmetry points, the gaps remain finite for all $\kk$, hence the projector to the central 10 bands per valley (equivalently, the central $20$ bands) in the $6 \times 6$ system is unambiguously defined and respects the symmetries of the system.

In an infinitely fine momentum grid band touchings appear at generic momenta, near the $\Gamma_M M_M$ line (or a symmetry-related line), and the band projector becomes ill-defined; these touchings are noticeable on the right column of Fig. \ref{fig8}(b) for a $96 \times 96$ grid. One can formally circumvent this problem by resorting to the projection onto the bands obtained from a Wannier orbital model that reproduces the symmetry representations and  dispersion \cite{carr19,po19}, and avoids the band touchings. The $6 \times 6$ projector approximates this regularized projector; discrepancies between the resulting effective theories (i.e. the $6 \times 6$ and regularized-projector theory) stem from the states near the band touchings, and are expected to be quantitatively minor. In particular, the main result of the nonflat bands in the $n_B=20$ effective theory will hold independently of the projector used. 

On another note, we have described above how our many-body projection algorithm takes into account interacting effects from the integrated-out states. Consequently, a non-projection (this is, a projection onto all bands) must return the original tight-binding system, $\mathcal{H}=H$, with the only difference that in $\mathcal{H}$ the density is measured with respect to $\rho_{0\boldsymbol{r}}$ (Eqs. (\ref{projhamiltoniansm}),(\ref{projhamiltoniansm2})) and in $H$ with respect to the uniform $1/2$ background (Eq. (\ref{hamiltoniansm})). $\mathcal{H}_0$ will then absorb the discrepancies in the definitions of the interaction. However, these differences are essentially negligible, as we can see in Fig. \ref{fig8}(c). This is consistent with the approximate relation $\rho_{0\boldsymbol{r}} = \sum_{\kk,n<0} |\langle  \boldsymbol{r} |  \kk n  \rangle|^2 \approx \frac{1}{2} \sum_{\kk n} |\langle  \boldsymbol{r} |  \kk n  \rangle|^2 = \frac{1}{2}$.

\begin{figure*}[h!]
    \includegraphics[width=.9\linewidth]{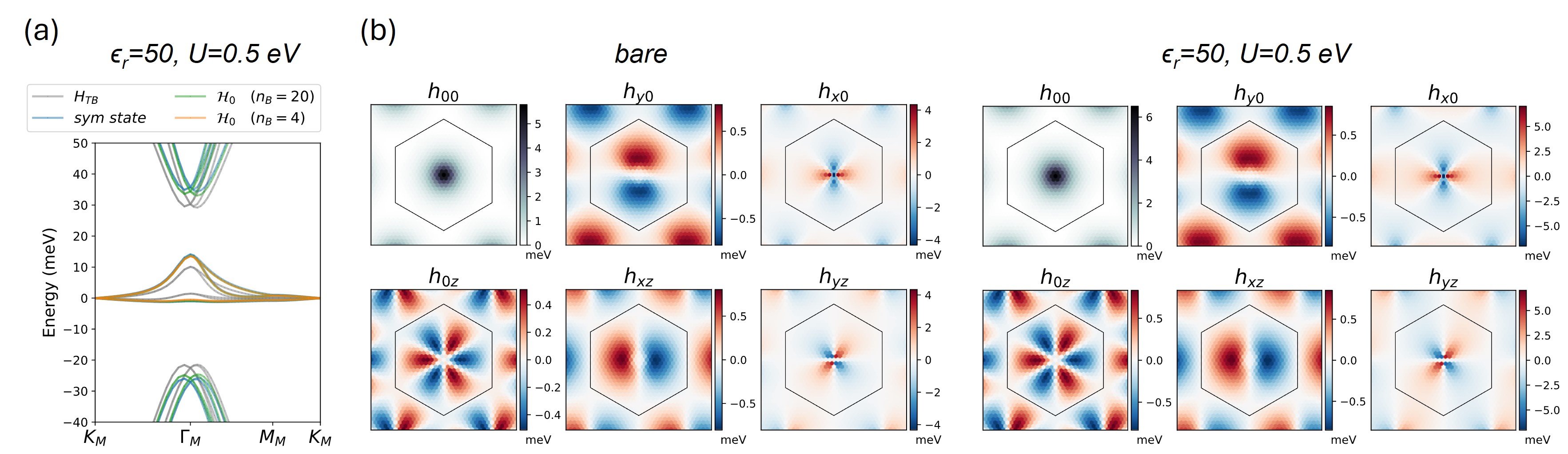}
    \caption{(a) Bare tight-binding, symmetric state and $\mathcal{H}_0$ ($n_B=4,20$) bands for coupling strength $\epsilon_r=50$, $U=0.5$ eV. Due to the weak coupling, the renormalized bands are only slightly wider. (b-c) The components of $h(\boldsymbol{k})$ of the flat-band theory ($n_B=4$) with (b) $\epsilon_r^{-1}=0$, $U=0$ eV (bare) and (c) $\epsilon_r=50$, $U=0.5$ eV.}
    \label{fig9}
\end{figure*}

In Fig. \ref{fig9}(a) we show the non-interacting (bare), the symmetric-state and the $\mathcal{H}_0$ (with $n_B=4,20$) band structures, for $\epsilon_r=50$, $U=0.5$ eV. In Fig. \ref{fig9}(b) we show the non-vanishing coefficients of $h(\kk)$ (Eq. \ref{hk}) for the non-interacting (bare) and $n_B=4$ system with $\epsilon_r=50$, $U=0.5$ eV. 

\clearpage

\section{Additional results: correlated states}\label{appg}

Assuming preserved moir\'e translations, the self-consistent states can be described by the one-particle density matrix,
    $P(\kk)_{\alpha \beta} = \frac{1}{2}\Big( \mathbb{1}_{\alpha \beta} + Q(\kk)_{\alpha \beta} \Big) = \langle  c^\dagger_{\kk \alpha}  c_{\kk \beta} \rangle,$
with the properties $P(\kk)^2 = P(\kk) = P(\kk)^\dagger$ and $\text{tr}(P(\kk)) =$ the number of occupied electrons at $\kk$. The indices $\alpha, \beta = (\tau \sigma s)$ denote the valley-sublattice-spin flavor and $\langle ...  \rangle$ denotes the expectation value. For spin-singlets and 4 occupied electrons at each $\kk$, $Q(\kk)$ can be decomposed into products of $\sigma$ and $\tau$ matrices with real coefficients, $Q(\kk) = \sum_{ij} \langle\sigma_i \tau_j (\kk) \rangle\sigma_i\tau_j$. Under a unitary transformation $\mathcal{U}_n(\varphi) = \text{exp}(i\varphi \mathcal{G}_n/2)$, with $\mathcal{G} = \sum_{\kk} c^\dagger_{\kk} G c_{\kk}$, the $Q(\kk)$ matrix transforms to $U^*_n(\varphi) Q(\kk) U^T_n(\varphi)$ with $U_n(\varphi) = \text{exp}(i\varphi G_n/2)$. 

We obtained self-consistent Hartree-Fock states in the flat-band ($n_B=4$) system on a $24 \times 24$ momentum grid. The wave functions of the flat bands were obtained by solving $H_\text{MF}$ (Eq. \ref{H_MF}) with the normal-state Fock matrix of the original $6\times6$ grid, on the $24 \times 24$ grid. The band structures were obtained on finer grids the using the same method with the self-consistent Fock matrix. 

\begin{figure*}[h!]
    \includegraphics[width=.9\linewidth]{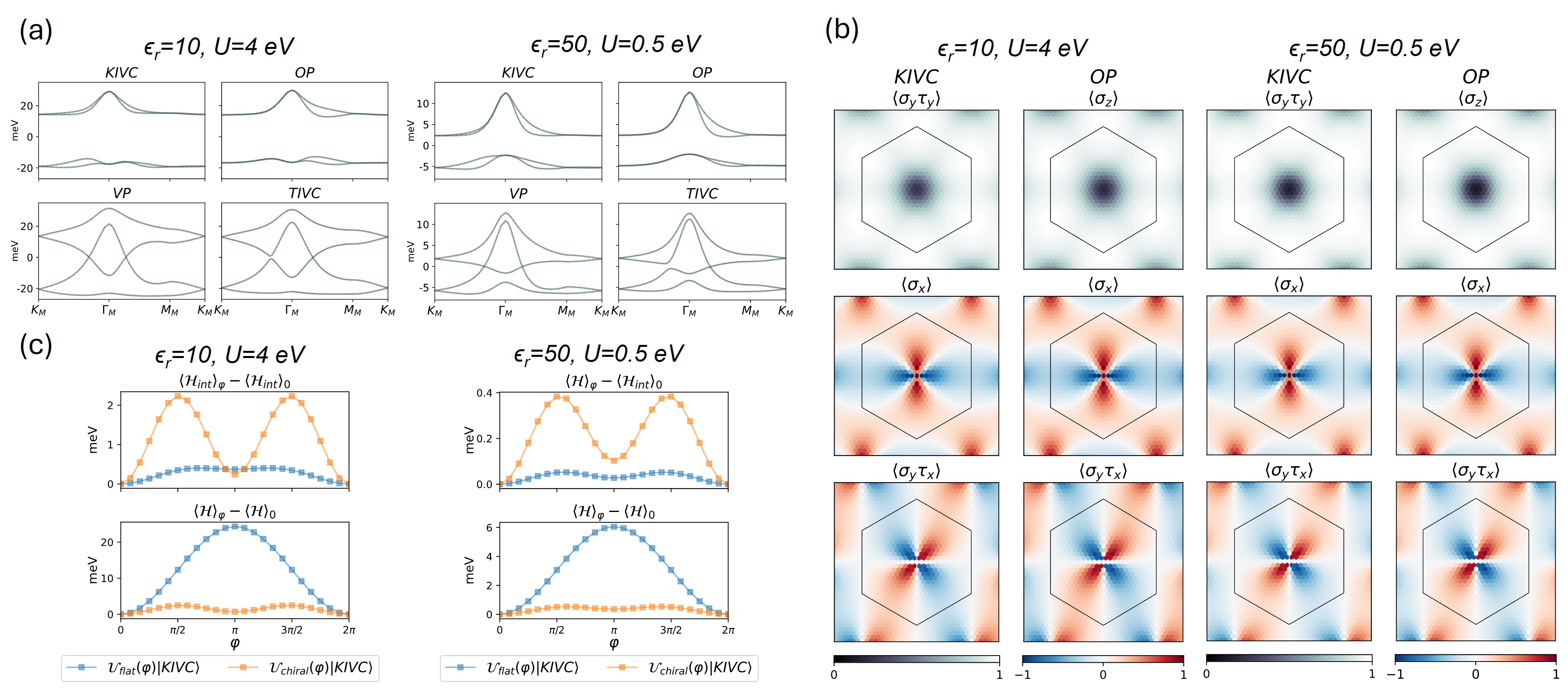}
    \caption{(a) Band structures of the Kramers intervalley coherent (KIVC), orbital polarized (OP), time-reversal invariant intervalley coherent (TIVC) and valley polarized (VP) states, for coupling strengths $\epsilon_r=10$, $U=4$ eV and $\epsilon_r=50$, $U=0.5$ eV. (b) Order parameters of KIVC and OP. (c) Expectation values of the energies of the rotated states $\mathcal{U}_{\text{flat}}(\varphi) |\text{KIVC}\rangle$ and $\mathcal{U}_{\text{chiral}}(\varphi) |\text{KIVC}\rangle$. The approximate chiral symmetry is manifest in (a-c) and the broken flat symmetry in (a).}
    \label{fig10}
\end{figure*}

In Fig. \ref{fig10} we study the KIVC (initial seed $Q(\kk) = \sigma_y\tau_y$ on the self-consistency loop), OP ($Q(\kk) = \sigma_z$), VP ($Q(\kk) =\tau_z$) and TIVC ($Q(\kk) = \sigma_x\tau_x$) states. Notice that, for the initial seeds, $|\text{KIVC} \rangle =  \text{exp}(i\pi/4 \ \mathcal{G}_{\text{chiral}})|\text{OP} \rangle$, $|\text{VP} \rangle =  \text{exp}(i\pi/4\ \mathcal{G}_{\text{chiral}})|\text{TIVC} \rangle$, $|\text{KIVC} \rangle =  \text{exp}(i\pi/4 \ \mathcal{G}_{\text{flat}})|\text{VP} \rangle$ and $|\text{TIVC} \rangle =  \text{exp}(i\pi/4 \ \mathcal{G}_{\text{flat}})|\text{OP} \rangle$.

In Fig. \ref{fig10}(a), KIVC and OP are insulating while VP and TIVC are gapless. This can be understood because KIVC and OP break $C_{2z}\mathcal{T}$ and open a gap at the $K$ points; while VP and TIVC preserve $C_{2z}\mathcal{T}$ and the Dirac cones do not gap out but split in energy. Given that the bandwidth of $\mathcal{H}_0$ is comparable to the interaction scale (the splitting at $K_M$), the system remains metallic.

This can also be understood in a different way. The properties $[\sigma_x\tau_x,\sigma_x]=[\tau_z,\sigma_x]=[\sigma_x\tau_x,\sigma_y\tau_z]=[\tau_z,\sigma_y\tau_z]= 0 $ and $P(\kk)^2 = P(\kk)$ disallow a smooth transition between the different orders ($\tau_z$, $\sigma_x \tau_x$ at the BZ boundaries and $\tau_x$, $\sigma_y \tau_z$ at the BZ center); this transition has to be discontinuous, i.e. through a band crossing. In the TIVC phase we find multiple additional order parameters like $\sigma_y$, $\sigma_x\tau_z$ and $\sigma_y\tau_x$ in a small region around the band crossing, and the transition is sharp, but smooth. The consequence is the band crossings become avoided crossing with small gaps. Moreover, the different $C_{2x}$ eigenvalues of the valence bands at $\Gamma_M$ and $M_M$ oblige a genuine crossing along the $\Gamma_M M_M$ line in the form of a Dirac cone \cite{angeli19}. On the other hand, the OP and KIVC states have $Q =\sigma_z$ and $Q=\sigma_y\tau_y$ respectively at the Brillouin zone boundaries. These matrices anticommute with $\sigma_x$ and $\sigma_y\tau_z$ and thus a continuous transition between the different order parameters is allowed; the self-consistent OP and KIVC states are insulators with a smooth $Q$ matrix (modulo the vortex at $\Gamma_M$ coming from the gauge choice).

In Fig. \ref{fig9}(b) we plot the main order parameters of KIVC and OP. The chiral symmetry relating the two phases is evident, as well as in Fig. \ref{fig9}(a) in the band structures of OP and KIVC, and VP and TIVC. 

In Fig. \ref{fig10}(c) we consider the self consistent KIVC state under flat and chiral transformations, $\mathcal{U}_{\text{flat}}(\varphi) |\text{KIVC}\rangle = \text{exp}(i \varphi/2 \ \mathcal{G}_{\text{flat}})  |\text{KIVC}\rangle $ and $\mathcal{U}_{\text{chiral}}(\varphi) |\text{KIVC}\rangle = \text{exp}(i \varphi/2 \ \mathcal{G}_{\text{chiral}})  |\text{KIVC}\rangle$. We plot the expectation value of the interaction ${\mathcal{H}}_\text{int}$ and the total energy ${\mathcal{H}}$ for varying $\varphi$. This provides an alternative confirmation that $\mathcal{H}_0$ $(={\mathcal{H}}-{\mathcal{H}}_{\text{int}})$ induces a strong $\mathcal{G}_{\text{flat}}$ breaking, while ${\mathcal{H}}_{\text{int}}$ is approximately $U(4) \times U(4)$ invariant. On the other hand, the amplitude of the ${\mathcal{H}}_{\text{int}}$ curves of the flat rotations being smaller than those of the chiral rotations was expected form the singular values of $C_{2z}\overline{P}$ being larger than the eigenvalues of $\overline{\sigma}_z$ (Fig \ref{fig6}(b) and Table \ref{tab3}).

\begin{figure*}[h!]
    \includegraphics[width=.9\linewidth]{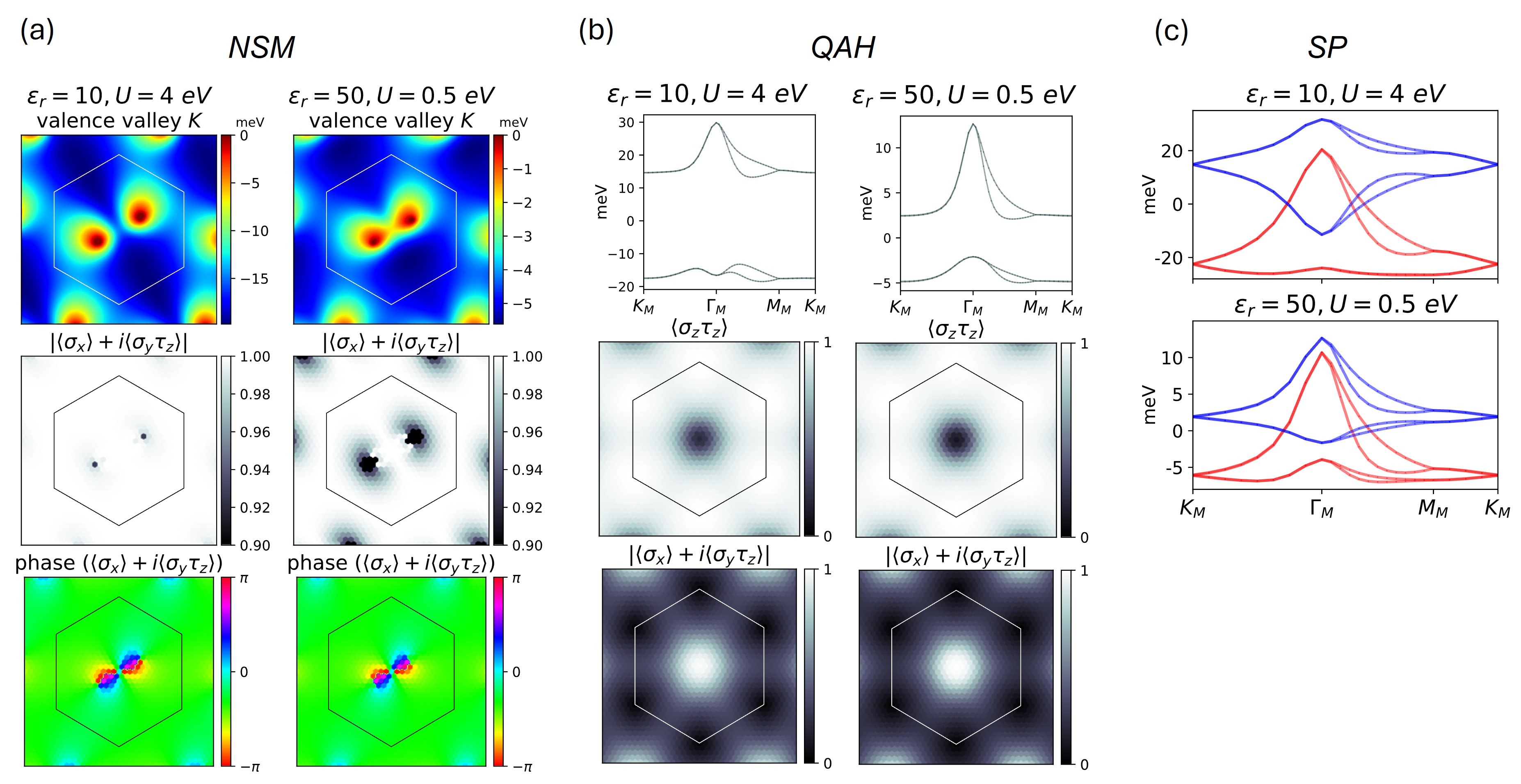}
    \caption{(a) Band structure of the valley $K$ valence band and order parameter of the nematic semimetal (NSM). (b) Band structure and order parameters of the Quantum Anomalous Hall (QAH) state. (c) Band structure of the spin polarized state (SP).}
    \label{fig11}
\end{figure*}

In Fig. \ref{fig11} we show the band structure and order parameters of the NSM, QAH and SP phases. The Dirac cones in NSM are visible in all panels, and the vortex at $\Gamma_M$ coming from the gauge choice is visible in the phase plot. The SP phase is metallic for the same reason as the VP.

% \clearpage
% \bibliography{bibliosup}

% \end{onecolumn}

\end{document}